\documentclass[11pt, letterpaper]{article}
\usepackage[margin=1in]{geometry}
\usepackage{mathtools}
\usepackage[linkcolor=blue]{hyperref}
\usepackage{amsmath}
\usepackage{amsthm}
\usepackage{amsfonts}
\usepackage{bbm}
\usepackage[capitalise]{cleveref}
\usepackage{algpseudocode}
\usepackage{algorithm}
\usepackage{bm}
\usepackage{setspace}
\usepackage{subcaption}
\usepackage[T1]{fontenc}
\usepackage{thmtools}
\usepackage{thm-restate}
\usepackage{enumitem}
\usepackage{xcolor}

\newcommand{\Exp}{\mathbb{E}}
\newcommand{\norm}[1]{\left\lVert#1\right\rVert}
\newcommand{\dualnorm}[1]{\norm{#1}^*}
\newcommand{\onenorm}[1]{\norm{#1}_1}
\newcommand{\inftynorm}[1]{\norm{#1}_{\infty}}
\newcommand{\topknorm}[1]{\norm{#1}_{\textnormal{top-$k$}}}
\newcommand{\orderednorm}[1]{\norm{#1}_{\beta}}
\newcommand{\dualorderednorm}[1]{\orderednorm{#1}^*}

\newcommand{\Rd}{\mathbb{R}^d}
\newcommand{\Rdpos}{\Rd_{\geq 0}}
\newcommand{\Rdstrictlypos}{\Rd_{> 0}}
\newcommand{\inp}[2]{\left\langle #1,#2 \right\rangle}
\newcommand{\sort}[1]{#1^\downarrow}

\newcommand{\probsimplex}{\Delta^d}
\newcommand{\dualspace}{Y}
\newcommand{\bigO}[1]{\mathcal{O}\left(#1\right)}
\newcommand{\bigOtilde}[1]{\tilde{\mathcal{O}}\left(#1\right)}
\newcommand{\Proj}{\mathrm{Proj}}
\newcommand{\LSE}{\mathrm{LSE}}
\newcommand{\GuessedOPT}{\Gamma}
\newcommand{\NormApproxError}{\delta}
\newcommand{\CustomerCurve}{\rho}

\DeclareMathOperator*{\OPT}{OPT}
\DeclareMathOperator*{\conv}{conv}
\DeclareMathOperator{\intervalratio}{ir}

\newtheorem{theorem}{Theorem}
\newtheorem{lemma}[theorem]{Lemma}
\newtheorem{corollary}[theorem]{Corollary}
\newtheorem{proposition}[theorem]{Proposition}
\newtheorem{definition}[theorem]{Definition}

\makeatletter
\let\@fnsymbol\@arabic
\makeatother

\title{An Efficient Algorithm for Minimizing Ordered Norms in Fractional Load Balancing\thanks{An extended abstract of this work appears in the Proceedings of IPCO 2026.}}
\date{}
\author{Daniel Blankenburg\thanks{
Research Institute for Discrete Mathematics and Hausdorff Center for Mathematics, University of Bonn, \texttt{\{blankenburg, ellerbrock, vygen\}@dm.uni-bonn.de}
}
\and 
Antonia Ellerbrock\footnotemark[2]
\and 
Thomas Kesselheim\thanks{
Institute of Computer Science and Hausdorff Center for Mathematics, University of Bonn, \texttt{thomas.kesselheim@uni-bonn.de}
}
\and 
Jens Vygen\footnotemark[2]
}

\begin{document}

\maketitle

\begin{abstract}
	We study the problem of minimizing an ordered norm of a load vector (indexed by a set of $d$ resources),
	where a finite number $n$ of customers $c$ contribute to the load of each resource by choosing a solution $x_c$ in a convex set $X_c \subseteq \mathbb{R}^d_{\geq 0}$;
	so we minimize $||\sum_{c}x_c||$ for some fixed ordered norm $||\cdot||$.
	We devise a randomized algorithm that computes a $(1+\varepsilon)$-approximate solution to this problem
	and makes, with high probability,
	$\bigO{(n+d) (\varepsilon^{-2}+\log\log d)\log (n+d)}$ calls to oracles that minimize linear functions (with non-negative coefficients) over $X_c$.

	While this has been known for the $\ell_{\infty}$ norm via the multiplicative weights update method, existing proof techniques do not extend to arbitrary ordered norms.

	Our algorithm uses a resource price mechanism that is motivated by the follow-the-regularized-leader paradigm, and is
	expressed by smooth approximations of ordered norms.
	We need and show that these have non-trivial stability properties, which may be of independent interest.
	For each customer, we define dynamic cost budgets, which evolve throughout the algorithm, to determine the allowed step sizes.
	This leads to non-uniform updates and may even reject certain oracle solutions.
	Using non-uniform sampling together with a martingale argument, we can guarantee sufficient expected progress in each iteration,
	and thus bound the total number of oracle calls with high probability.
\end{abstract}

\setcounter{page}{0}
\newpage
\section{Introduction}

Many fractional load balancing problems can be understood as multiple customers sharing multiple resources with the goal of minimizing the maximum resource consumption.
A simple example is fractional scheduling, where jobs have to be assigned to machines to minimize the makespan.
Similarly, fractional packing~\cite{plotkin1995fast} can be formulated as finding a probability distribution $x$ that minimizes the $\ell_{\infty}$ norm of $Ax$ for some non-negative matrix $A$.
Another such problem is maximum concurrent flow, where we have to send flows in a graph from source nodes to respective sink nodes such that all demands are fulfilled and the highest congestion of an edge is minimized.
This problem can be generalized in multiple ways.
For example, a highly relevant problem for VLSI routing is fractional Steiner tree packing for given terminal sets to minimize the worst congestion~\cite{hu2001survey, jain2003packing, muller2011faster}.
In all these problems, the standard goal is to minimize the maximum utilization/load/congestion of any resource,
which can also be understood as minimizing the $\ell_\infty$ norm of the load vector.

However, minimizing the $\ell_\infty$ norm is not very robust.
In VLSI routing, due to modeling issues, a few edges with very high congestion can often not be avoided.
Therefore, design engineers typically use measures such as the average congestion of the worst $x \%$ of resources
(i.e., a top-$k$ norm) or linear combinations thereof~\cite{wei2014techniques}.
This corresponds to an ordered norm applied to the load vector.

In this paper, we consider the problem of minimizing an ordered norm in fractional load balancing problems.
We devise a fast algorithm that computes a $(1 + \varepsilon)$-approximate solution for any given $\varepsilon>0$.

\subsection{Problem Statement}

We assume that there is a set $C$ of $n$ ``customers'' which correspond to jobs, commodities, or requests in the underlying problem.
For each customer $c \in C$, there are multiple ways to serve them.
We identify these ways with their resource utilization by assuming a closed convex set $X_c \subseteq \Rdpos$ per customer, 
where a vector $x_c \in X_c$ denotes what fraction of each of the available $d$ resources is used.
The set of feasible solutions to our overall problem is $X = \sum_{c \in C} X_c$.
Our objective is to find an $x \in X$ that (approximately) minimizes $\orderednorm{x}$, where $\orderednorm{\cdot}$ is an ordered norm, i.e.,
\begin{equation}
	\min_{x \in X} \ \orderednorm{x} \enspace . \label{eq:problem_definition}
\end{equation}
An ordered norm is defined by $\orderednorm{x} = \sum_{i = 1}^d \beta_i (\sort{x})_i$ for $x\in\Rdpos$, 
for some non-increasing weight vector $\beta \in \Rdpos$, where $(x^{\downarrow})_i$ denotes the $i$-th largest entry in $x$.
Recovering the $\ell_1$ norm when all weights are equal and the $\ell_{\infty}$ norm when only the first weight $\beta_1$ is non-zero,
ordered norms are a natural choice to interpolate between minimizing the average and the maximum resource usage.

We assume that the sets $X_c$ are not given explicitly but by access to an oracle minimizing a linear function (with non-negative coefficients) over the fixed set $X_c$, an assumption that is prevalent in theory and in practice~\cite{braun2025conditional,jaggi2013revisiting}.
In the special case of fractional packing, this means that $A$ can have exponentially many columns and is not given explicitly;
instead, we have an oracle that returns one of these columns (minimizing the inner product with a given price vector).

A well-known special case is the concurrent flow problem. Here, the customers correspond to the commodities.
If customer $c$ can be served by sending one unit of flow from $s$ to $t$ in a unit-capacity graph, 
then $d$ is the number of edges and $X_c$ is the convex hull of incidence vectors of edge sets of $s$-$t$-paths.
In this case, the linear minimization subproblem corresponds to finding a shortest $s$-$t$-path.
In the maximum concurrent flow problem, we aim at minimizing the $\ell_{\infty}$ norm over~$X$.
We will generalize this to arbitrary ordered norms; see \cref{cor:concurrentflow}.

\subsection{Our Main Result}

Our main result is the following. 
Recall that $n=|C|$ denotes the number of customers, and $d$ denotes the number of resources (so $X=\sum_{c \in C}X_c\subseteq\Rdpos$).
\begin{theorem} \label{thm:main}
	There is a randomized algorithm that computes, for any given $\varepsilon>0$,
	a $(1+\varepsilon)$-approximate solution to~\eqref{eq:problem_definition}
	and makes, with high probability,
	$\bigO{(n+d) (\varepsilon^{-2}+\log\log d)\log (n+d)}$ calls to oracles that minimize linear functions with non-negative coefficients over $X_c$ for $c \in C$.
	The additional time per oracle call is $\bigO{d \log d}$.
\end{theorem}

This also works with $\tau$-approximate linear minimization oracles (for any $\tau\ge 1$),
and leads to a $\tau(1+\varepsilon)$-approximate solution to \eqref{eq:problem_definition}.

The dependence on $\varepsilon$ is best possible, as shown in~\cite{klein2015number}.
Further, our running time matches the best known running time for the $\ell_{\infty}$ norm~\cite{muller2011faster, blankenburg2022resource} up to $\log$-factors.

As an application, we can minimize arbitrary ordered norms rather than just the $\ell_{\infty}$ norm in the concurrent flow problem.
The ordered norm is applied to the congestion vector of a multi-commodity flow. The congestion vector 
has an entry for each edge, which is the total flow along this edge divided by the edge's capacity.

\begin{corollary}
	\label{cor:concurrentflow}
Given a multi-commodity flow instance with $v$ vertices, $d$ edges with positive real capacities, and $k$ commodities with positive real demands, 
and any ordered norm $\orderednorm{\cdot}$,
we can compute a multi-commodity flow with congestion vector $x$ such that $\orderednorm{x}$ is minimum up to a factor $1+\varepsilon$, in
$\bigO{d(d+\min\{k,v\})(\varepsilon^{-2}+\log\log d)\log^2 d}$
time. 
\end{corollary}

\begin{proof}
Define a customer for each vertex that is the source of at least one source-sink pair. We have $n\le\min\{k,v\}$ customers, 
and we may assume $v\le d$.
For a customer (source) $c$, let $\text{sinks}(c)$ denote the set of corresponding sinks; then 
\[
X_c \ = \ \conv\Biggl\{ \Biggl(  \frac{1}{u_e} \sum_{s\in\text{sinks}(c):e\in P_s} d_{cs} \Biggr)_{e\in E} : P_s \text{ path from source $c$ to sink $s$ for }s\in\text{sinks}(c) \Biggr\} \enspace,
\]
where $E$ is the set of edges, $u_e>0$ is the capacity of edge $e$, and $d_{cs}>0$ is the demand of the commodity $c\rightarrow s$. 
To minimize a linear function over $X_c$, we apply Dijkstra's algorithm to compute
a shortest path from source $c$ to each sink in $\text{sinks}(c)$ in total time $\bigO{d\log d}$. The result follows from \cref{thm:main}.
\end{proof}

Since our main result extends to oracles that only approximately minimize linear functions over $X_c$ for $c\in C$ (see \cref{section:approximate_oracles}),
we can also approximate ordered norms for fractional Steiner tree packing and other resource sharing problems of crucial relevance in chip design~\cite{muller2011faster,wei2014techniques,held2017global,daboul2018provably,daboul2023global}.

\subsection{A Simple Algorithm and its Limitations}
\label{sec:simplealganditslimitations} 

Our algorithm is based on a well-known algorithm template, exploiting duality:
Every norm can also be written as $\norm{x} = \max_{y \in Y} \langle x, y \rangle$ for $x \in \Rdpos$, where the dual space $Y$ is the set of non-negative non-dominated vectors with dual norm equal to $1$.
Therefore, solving problem~\eqref{eq:problem_definition} is equivalent to finding an equilibrium strategy of the minimization player in the zero-sum game $\min_{x \in X} \max_{y \in Y} \langle x, y \rangle = \max_{y \in Y} \min_{x \in X} \langle x, y \rangle$.

In order to obtain a solution to \eqref{eq:problem_definition}, one determines $x$ and $y$ in an iterative way, while interpreting the vector $y$ as positive resource prices. The algorithm alternates between calls to a minimization oracle, which determines the cheapest solution $x$ based on the current prices $y$, and calls to some price update mechanism, which updates $y$ based on all previous solutions $x$. After a fixed number of iterations, the average of all encountered solutions $x$ is returned.
Common price update mechanisms include Mirror Descent, Follow-the-Regularized-Leader, or Follow-the-Perturbed-Leader~\cite{Lattimore_Szepesvari_2020, shalev2012online}; see \cref{sec:related_work} for an overview of existing approaches. 

In particular, in case of the $\ell_\infty$ norm, one can use the Multiplicative Weight Update or Hedge algorithm \cite{arora2012multiplicative, freund1997decision} to determine the prices; here, the dual space is the probability simplex.
Moving to more general norms and thus dual spaces $Y$, one can show, using Follow-the-Regularized-Leader to update the prices,
that $\bigO{\varepsilon^{-2} \log d }$ oracle calls suffice to obtain a guarantee of $\exp(\varepsilon) \OPT + \varepsilon$ if $X \subseteq [0,1]^d$ (see \cref{sec:single_customer_continuous}), where $\OPT$ denotes the optimum solution value of \eqref{eq:problem_definition}.
Similar guarantees follow with the standard Frank--Wolfe algorithm \cite{jaggi2013revisiting} with a suitable smooth approximation of the given norm (for example, the one we discuss in \cref{subsec:norm_approx}).

However, these results are not quite satisfactory for two reasons.
First of all, given the additive error, the bound is only meaningful if $\OPT$ is sufficiently large. This is a strong assumption when $X \subseteq [0,1]^d$.
Removing the assumption $X \subseteq [0,1]^d$ will generally increase the running time by a factor proportional to the squared $\ell_{\infty}$ diameter of $X$.

Secondly, and even more severely, in every iteration, the algorithm needs to determine an $x \in X$ that minimizes $\langle x, y \rangle$ given the current prices $y$.

In principle, a linear minimization oracle over $X$ can be implemented by querying the oracle for each set $X_c$ and summing the resulting vectors.
Thus, a single oracle call over $X$ requires $n$ oracle calls to the individual customer oracles.
Consequently, any method that treats $X$ as a single black-box domain and performs $k$ oracle calls over $X$ incurs $\Omega(nk)$ customer-oracle calls.

Even for a single customer, dimension-dependent lower bounds arise.
For example, consider minimizing the $\ell_{\infty}$ norm over the simplex $\probsimplex = \{ x \in [0,1]^d : \onenorm{x} = 1 \}$.

All extreme points of $\probsimplex$ are unit vectors with a single entry equal to $1$ and all other entries equal to zero.
Any $(1+\varepsilon)$-approximate solution must be positive on $\Omega(d)$ entries.
Consequently, any such solution requires $\Omega(d)$ extreme points in its convex decomposition.
Since each linear minimization oracle call over the simplex returns a single extreme point, it follows that $\Omega(d)$ oracle calls are necessary even in this simple case.

Combining these two observations shows that approaches operating only on the set $X$ cannot avoid a number of oracle calls of order $\Omega(n\cdot d)$ in general.
Our goal, by contrast, is to obtain a number of oracle calls that is near-linear in $n+d$ (and at most quadratic in $\varepsilon^{-1}$).
Achieving such bounds therefore requires exploiting the multi-customer structure more carefully rather than treating $X$ as a black-box set.

More generally, this lower-bound argument can be extended by considering an instance in which the feasible load vectors of one customer are given by the simplex, while all other customers have only the zero vector as a feasible solution.
In this case the problem essentially reduces to the simplex example above, yet any algorithm that queries all customer oracles a similar number of times must still perform $\Omega(d)$ oracle calls for the simplex customer.
Since there are $n$ customers, this implies a total of $\Omega(n\cdot d)$ customer-oracle calls.

Consequently, algorithms that treat all customers symmetrically, such as methods that query each customer the same number of times per phase or sample customers uniformly at random in every iteration, cannot overcome the $\Omega(n\cdot d)$ barrier.
Breaking this barrier therefore requires algorithms that exploit the multi-customer structure in a non-uniform way.

\subsection{Our Techniques}
\label{subsec:our_techniques}

In order to prove \cref{thm:main}, we design a new algorithm that is inspired by the simple one sketched in the previous section but requires several new ideas.

Our price update strategy is motivated by Follow-the-Regularized-Leader with the negative entropy $F(y) = \sum_{i=1}^d y_i \ln y_i$ as regularizer.
The prices can be expressed equivalently as the gradients of the convex conjugate of $F$ on the dual space $Y$, given by $F^*(x) = \max_{y \in Y} (\inp{y}{x} - F(y))$.
This connection is known as Danskin's theorem~\cite{danskin1966theory}.
This convex conjugate coincides with the \textsc{LogSumExp} function for the $\ell_{\infty}$ norm.

In this work, it is more convenient to use the language of norm approximations.
Indeed, the convex conjugate of the negative entropy gives rise to a norm approximation that satisfies the properties of~\cref{thm:norm_approximation} (see~\cref{sec:construction_norm_approximation}).
This result could be of independent interest in future work.

\begin{restatable}[Ordered norm approximation]{theorem}{normapproximation}
	\label{thm:norm_approximation}
	For every ordered norm $\orderednorm{\cdot}$ on $\Rd$ and every $\eta > 0$, there is a differentiable convex function $\Psi$ fulfilling $\orderednorm{x} \leq \Psi(x) \leq \orderednorm{x} + \frac{\ln d}{\eta}$ and $\nabla \Psi(x+y) \ \leq \ \exp \left( \eta \inftynorm{y} \right) \cdot \nabla \Psi(x)$ for all $x, y \in \mathbb{R}^d_{\geq 0}$.
	Moreover, $\Psi$ and its gradient can be evaluated in $\bigO{d \log d}$ time.
\end{restatable}

We prove this in~\cref{sec:construction_norm_approximation,sec:proof_contraction_property}.
The property of $\nabla \Psi$ (bounded gradient increase) follows from 
structural properties of the generalized relative entropy projection to the dual space~$Y$, in particular a contraction property that we prove for ordered norms.
The bounds on $\Psi$ and $\nabla\Psi$ are best possible; see \cref{prop:bestpossiblenormapx}.

The approximation and stability properties in \cref{thm:norm_approximation} are similar to the ones in \cite{molinaro2017online,kesselheim2022online} but there are subtle differences. In contrast to~\cite{kesselheim2022online}, we bound the increase of the gradient while in \cite{kesselheim2022online} there is only a bound on the decrease. Besides, our bound on the gradient increase depends on the $\ell_\infty$ norm of the change, which is generally a looser bound.
The norm approximations proposed in~\cite{molinaro2017online} require the same upper bound on the maximum gradient increase as in~\cref{thm:norm_approximation}, but additionally impose a lower bound on the maximum gradient decrease, which we do not need.
Further, these approximations were constructed only for $\ell_p$ norms.

If we had a norm approximation $\Psi$ as in \cref{thm:norm_approximation} (cf.~\cref{def:norm_approximation}) for more general classes of norms, which additionally satisfy the scaling assumption $\norm{1^d}=1$ and \mbox{$\frac{\onenorm{\cdot}}{d} \leq \norm{\cdot}$}, then \cref{thm:main} would also hold for those.

The norm approximation properties of \cref{thm:norm_approximation} imply \emph{weak duality} (see \cref{lem:gradients_dual_norm_bounded_by_1,cor:inp_bounded_by_OPT}):
\[
	\min_{s \in X} \inp{\nabla \Psi(x)}{s} \ \leq \ \OPT \qquad \forall x \in \Rdpos \enspace.
\]

Our core algorithm (\cref{alg:general_ordered_norm_minimization_multiple_customers_continuous}) collects, for each customer, solutions of total weight $T \in \mathbb{N}$ and takes the weighted sum. 
While the total weight of the solutions per customer is $T$, the number of oracle calls can be significantly smaller or larger than $T$.
Since the step sizes of our algorithm are non-uniform, it is convenient to parametrize the algorithm according to the total solution weight that is collected.
We interpret the evolving partial solution of our algorithm as a piecewise linear curve $S:[0, nT] \rightarrow \Rdpos$.
The curve starts at $S(0) = 0$ and ends in $S(nT)$.
The output of our algorithm will be $\frac{1}{T} S(nT)$.
We build $S$ by successively adding oracle solutions with a carefully chosen weight.
That is, at a time point $t \in [0,nT]$, we query the oracle of a customer $c$ with the prices $\nabla \Psi(S(t))$, and
add its returned solution with weight $\xi \geq 0$, i.e., define $S(t+\xi') = S(t) + \xi' s$ for all $0\leq \xi' \leq \xi$.
We use the gradient theorem for line integrals to bound the cost of the final solution
\[
	\Psi(S(nT)) \ = \ \Psi(S(0)) \ + \ \int_{0}^{nT} \inp{\nabla \Psi(S(t))}{S'(t)} \ \text{d}t \enspace .
\]
This is a continuous analogue to regret bounds of Follow-the-Regularized-Leader, and shares similarities with a continuous time analysis of the multiplicative weight update method as in~\cite{chekuri2015multiplicative}.
The derivative $S'(t)$ is a direction of the curve $S$ and thus always corresponds to the last added oracle solution (up to a finite number of points where $S$ is not differentiable).
We may interpret the term $ \int_{0}^{nT} \inp{\nabla \Psi(S(t))}{S'(t)} \, \text{d}t$ as the "cost incurred along the way".
Instead of the usual viewpoint, where we collect a solution $s$ with coefficient $\xi$ and incur a cost of $\inp{y}{\xi s}$ for the fixed price $y$ that was given as an input to the oracle,
we pay for the solution $s$ under the evolving costs $\nabla \Psi(S(t))$ on an interval of length $\xi$.
To analyze these costs, the bounded increase of $\nabla \Psi$ as well as weak duality are crucial.
In~\cref{sec:single_customer_continuous}, we provide an analysis of the standard case $n = 1$ and $X\subseteq [0,1]^d$ that illustrates this technique.
This is significantly simpler than the general case, since it does not work with multiple customers and unbounded domains.

The algorithm for the $\ell_{\infty}$ norm by~\cite{muller2011faster} proceeds in $T$ phases, where, in every phase, oracle
solutions of total weight one are collected for each customer.
The analysis crucially relies on the fact that $\nabla \Psi$ remains stable for the entire phase, and the incurred cost
in a phase can be bounded by approximately $\OPT$.
For general ordered norms, stability of $\nabla \Psi$ cannot be guaranteed for an entire phase, and this proof technique does not work.
Thus, our algorithm does not proceed in phases, and requires a completely different proof.

Indeed, our new proof technique can be considered as one of the main technical contributions of our work.
We use a \emph{dynamic cost budget} $\sigma:[0,nT) \rightarrow [0,1]^n$, with $\sum_{c \in C} \sigma_c(t) = 1$ at all times $t \in [0,nT)$,
which represents the fraction of $\OPT$ that each customer is allowed to cost at time $t$.
We call customer $c$ \emph{good} at time $t$ if their cost is not much larger than their cost budget.
Assuming that we know $\OPT$, this means:
\[
	\min_{s \in X_c} \inp{\nabla \Psi(S(t))}{s} \ \leq \ \exp(2\eta) \cdot \sigma_c(t) \cdot \OPT \enspace .
\]
Our algorithm accepts only oracle solutions of good customers.
Weak duality implies that there always exists a good customer.
To ensure feasibility of the final solution $\frac{1}{T}S(nT)$, solutions of total weight $T$ must be collected for every customer.
We encode this directly in the cost budget.
If, at time $t$, solutions of total weight $T$ have already been collected for customer $c$, we have $\sigma_c(t) = 0$.
We call such customers \emph{inactive} at time $t$, and \emph{active} otherwise.
Only active customers can be good.
With this strategy, we can bound the total cost by
\[ 
\int_{0}^{nT} \inp{\nabla \Psi(S(t))}{S'(t)} \ \text{d}t \ 
\leq \ \exp(2\eta) \left(\int_{0}^{nT} \sigma_{c(t)}(t) \ \text{d}t\right) \OPT \enspace ,
\]
where $c(t)$ is the customer whose oracle solution was chosen last before time $t$.
We show that our dynamic cost budget satisfies $\int_{0}^{nT} \sigma_{c(t)}(t) \ \text{d}t \leq T + \frac{\ln n}{\eta}$ for every realization of our randomized algorithm,
which is a key ingredient of our proof.
Further, our cost budget does not decrease too rapidly:
For every \mbox{$\xi > 0$} such that customer $c$ is active at time $t+\xi$, we have
\[
	\sigma_c(t+\xi) \ \geq \ \exp(-\eta \xi) \ \sigma_c(t) \enspace .
\]
Together with the bounded gradient increase of $\nabla \Psi$ (cf.\ \cref{thm:norm_approximation}), 
this allows us to take sufficiently large steps $\xi$, while guaranteeing that the chosen customer remains good on the entire interval $[t,t+\xi)$.
We choose the cost budget as the softmax of the remaining solution weight that needs to be collected for each active customer.

Finding a good customer may require $n$ oracle calls in general.
Another major technical contribution is the reduction of the number of oracle calls through non-uniform sampling.
We emphasize that the performance guarantee of our algorithm is deterministic, and only the running time is affected by randomization.
In every iteration, we sample a customer $c$ according to the cost budget $\sigma$ and take a maximal step $\xi$ that guarantees that customer $c$ remains good for the entire step length.
By calling the oracle for $c$, we observe the true fraction $q_c(t) \in [0,1]$ that the solution for customer~$c$ costs relative to $\OPT$.
The step length $\xi$ depends on the ratio $\sigma_c(t) / q_c(t)$ and a width reduction technique similar to~\cite{garg2007faster}.
It may happen that $\xi = 0$, but we use a martingale argument to show that, in expectation, $\xi$ is sufficiently large.

\sloppy
Our core algorithm assumes $\OPT \leq 1$, and returns a solution of value \mbox{$\varepsilon + \exp(\varepsilon)$}
after \mbox{$\bigO{\varepsilon^{-2}(n+d) \log (d+n)}$} oracle calls with high probability.
To bound the number of oracle calls in the worst case, and to guarantee $\OPT \leq 1$,
we design a stronger variant, \cref{alg:main_algorithm_with_guaranteed_termination}.
This is guaranteed to terminate and either correctly decides that $\OPT > 1$ or computes a solution of value $\varepsilon + \exp(\varepsilon)$.
Overall, we apply this to scaled instances and embed this into a two-stage binary search to determine $\OPT$ approximately.

\subsection{Related Work}

Minimizing norms in load balancing problems has been studied extensively in the literature.
Our work builds upon and brings together different lines of work,
including online and offline techniques.
The following gives a brief overview, whereas a detailed one is provided in \cref{sec:related_work}.

By formulating our problem as a zero-sum game, one can apply no-regret algorithms for online linear optimization, 
using price update mechanisms such as Follow-the-Regularized-Leader, Mirror Descent, or Follow-the-Perturbed-Leader.
For a single customer and several specific types of norms, our problem is well-solved~\cite{helmbold2009learning,herbster2001tracking,koolen2010hedging}.
This line of work primarily emphasizes price updates and can be interpreted as taking a \textit{dual} perspective.
In contrast, first-order methods that operate directly on the set of feasible solutions offer a \textit{primal} perspective.
When linear minimization oracles are available, our problem can be tackled using variants of conditional gradient descent~\cite{jaggi2013revisiting,lacoste2013block} applied to a differentiable approximation of the given norm.
An extensive overview of many of these techniques, together with a unified framework describing their connections, is provided in~\cite{wang2024no}.
However, this survey does not consider the multi-customer setting studied here.

Naturally, the primal and dual perspective are closely connected: gradients of smooth norm approximations define dual vectors -- a connection noted previously by many authors~\cite{abernethy2014online, kesselheim2022online}.
Our work is no exception; it seeks to further explore the interplay between primal and dual viewpoints, drawing inspiration from both and implicitly applying techniques from online optimization to an offline setting.

A separate line of earlier work,
and at the same time our original motivation for the problem, 
comes from multi-commodity flow~\cite{fleischer2000approximating,garg2007faster} and resource sharing under the $\ell_{\infty}$ norm~\cite{grigoriadis1994fast,muller2011faster},
which also yields width reduction strategies and methods for handling multiple customers.
For the special case of multi-commodity flow in the $\ell_\infty$ case,
\cite{garg2007faster} and \cite{fleischer2000approximating} combine a width reduction technique with a round robin scheme
and obtain an algorithm that requires only $\bigOtilde{ \varepsilon^{-2}(n+d) }$ calls to single-customer oracles.
\cite{muller2011faster} use a similar approach for the general $\ell_{\infty}$ case.
The underlying reason for these positive results is that for the $\ell_{\infty}$ norm, prices have a simple closed-form representation, and satisfy strong stability properties~\cite{kesselheim2022online}, which do not hold for general ordered norms.
We generalize these techniques to ordered norms.

While minimization of $\ell_\infty$ and $\ell_p$ norms has been studied for decades, ordered norms and other (monotone symmetric) norms have only been studied quite recently. 
One line of research~\cite{chakrabarty2019approximation,chakrabarty2019simpler,ibrahimpur2021minimum} considers offline \emph{integral} load balancing problems.
The techniques are quite different as they mostly focus on rounding fractional solutions of LP relaxations and only obtain constant-factor approximations. 
Closer in spirit are recent works on online problems~\cite{kesselheim2022online,kesselheim2024supermodular} that also use norm approximations.
However, the obtained approximation ratios are only poly-logarithmic.

\section{Approximating Ordered Norms}
\label{sec:basics}

\subsection{Ordered Norms}

To define ordered norms,
let $\sort{x}$ arise from $x\in\Rd$ by sorting the entries in non-increasing order.
We call $x$ \emph{sorted} if $x=\sort{x}$.
Further, we denote $|x| \coloneqq (|x_1|, \dots, |x_d|)$.

\emph{Ordered norms} (on $\Rd$) are all norms that can be written as
\[
\orderednorm{x} \ \coloneqq \ \sum_{i=1}^d \beta_i \left(\sort{|x|} \right)_i \qquad \text{ for all } x \in \Rd 
\]
for some sorted vector $\beta \in \Rdpos$.

Note that every ordered norm is monotone and symmetric.
In the following, we will assume without loss of generality that our ordered norms are \emph{normalized} in the sense that $\onenorm{\beta} = 1$.
This gives the useful property that $\orderednorm{1^d} = 1$, where $1^d=(1,\ldots,1)$ denotes the all-one vector.
Similarly, we write $0^d = (0,\dots, 0)$.

Simple examples of ordered norms are $\ell_{\infty}$, described by $\beta = (1,0,\dots,0)$, and $\ell_1$, described by $\beta = 1^d$. 
The $\ell_1$ norm is not normalized and we would thus consider $\frac{1}{d} \cdot \ell_1$, described by $\beta = \frac{1}{d} \cdot 1^d$.
Both $\ell_{\infty}$ and $\ell_1$ are also \emph{top-$k$ norms}, a subclass of ordered norms. 
After normalizing, the top-$k$ norm is given by $\orderednorm{\cdot}$ 
with $\beta_i=\frac{1}{k}$ for $i \in \{1, \dots , k\}$ and $\beta_i = 0$ for $i > k$.
The ordered norms are precisely the weighted sums 
$\sum_{k=1}^d \gamma_k \topknorm{\cdot}$ of top-$k$ norms with coefficients $\gamma_k \geq 0$.

The \textit{dual norm} of an ordered norm $\orderednorm{\cdot}$ is defined as
\[
	\dualorderednorm{y} \ \coloneqq \ \max \left\{ \inp{x}{y} \ : \ x \in \Rdpos \ , \ \orderednorm{x} \leq 1 \right\} \enspace .
\]

Ordered norms have the useful property that they can be bounded by the $\ell_1$-norm, which we use in several places in our analysis:
\begin{proposition}
\label{prop:l1_bound}
For every (normalized) ordered norm $\orderednorm{\cdot}$ on $\Rd$ and every $x \in \Rdpos$, 
\[
\frac{\onenorm{x}}{d} \ \leq \ \orderednorm{x} \ \leq \ \inftynorm{x} \enspace.
\]
\end{proposition}
\begin{proof}
Note that $\dualorderednorm{\frac{1}{d}1^d} =1$.
Therefore, we have $\orderednorm{x} \geq \inp{\frac{1}{d}1^d}{x} = \frac{\onenorm{x}}{d}$.
Now, since $\orderednorm{1^d} =1$,
\[
\orderednorm{x} \ \leq \ \orderednorm{\max_{i=1,\dots, d} x_i \cdot 1^d } \ = \ \inftynorm{x} \enspace . \qedhere
\]
\end{proof}

\subsection{Norm Approximation}
\label{subsec:norm_approx}
In this section, we define a \textit{norm approximation}, similar to \cite{kesselheim2022online}, which is central in the analysis of our algorithms.
The prices (dual vectors) during our algorithm will be the gradients of such a norm approximation.

\begin{definition}[Norm approximation]
	\label{def:norm_approximation}
Let $\NormApproxError, \eta \in \mathbb{R}_{> 0}$. A differentiable convex function $\Psi: \Rdpos \rightarrow \mathbb{R}_{\geq 0}$ is a \emph{$(\NormApproxError, \eta)$-approximation} of an ordered norm $\orderednorm{\cdot}$ on $\Rd$ if the following two properties hold:

\begin{enumerate}
\item \emph{Norm approximation} with absolute error $\NormApproxError$:
\[
\forall x \in \mathbb{R}^d_{\geq 0}: \quad \orderednorm{x} \ \leq \ \Psi(x) \ \leq \ \orderednorm{x} + \NormApproxError
\]
\item \emph{Bounded gradient increase} with parameter $\eta$:
\[
\forall x,y \in \mathbb{R}^d_{\geq 0}: \quad \nabla \Psi(x+y) \ \leq \ \exp \left( \eta \inftynorm{y} \right) \cdot \nabla \Psi(x)
\]
\end{enumerate}
\end{definition}
Note that given a $(\NormApproxError, \eta)$-approximation $\Psi$, one can construct a $\left( \frac{\NormApproxError}{s}, s \eta \right)$-approximation for every $s  >0$ by considering $x \mapsto \frac{1}{s} \Psi(sx)$.
We now show that every norm approximation $\Psi$ is monotone and $\dualorderednorm{\nabla \Psi} \leq 1$:

\begin{lemma}
\label{lem:gradients_dual_norm_bounded_by_1}
Let $\Psi$ be a $( \NormApproxError, \eta)$-approximation of an ordered norm $\orderednorm{\cdot}$ on $\Rd$.
Then, $\nabla \Psi(x) > 0^d$ and $\dualorderednorm{\nabla \Psi(x)} \leq 1$ for all $x \in \Rdpos$.
\end{lemma}

\begin{proof}
To prove positivity of the gradients, assume that there exists an $x \in \Rdpos$ with $(\nabla \Psi(x))_i \leq 0$ for some $i \in \{1,\dots,d\}$.
Bounded gradient increase of $\Psi$ implies that $(\nabla \Psi(x+K \mathbbm{1}_i))_i \leq 0$ for every $K > 0$, where $\mathbbm{1}_i$ is the $i$-th unit vector, and thus $\Psi(x+K\mathbbm{1}_i) \leq \Psi(x)$.
This is a contradiction to the norm approximation property of $\Psi$ for $K$ large enough, because $K \beta_1 \leq \Psi(x+K\mathbbm{1}_i)$.

To prove that the dual norm of the gradients is bounded by $1$, assume that there is an $x \in \Rdpos$ with $\dualorderednorm{\nabla \Psi(x)} = \omega > 1$,
and let $y \in \Rdpos$ with $\orderednorm{y} = 1$
such that $\inp{\nabla \Psi(x)}{y} = \omega$. Let $K > \frac{\orderednorm{x} + \NormApproxError}{\omega - 1}$.
By convexity and non-negativity of $\Psi$, we have
\begin{equation*}
	\Psi(x + Ky) \ \geq \ \Psi(x) +  \inp{\nabla \Psi(x)}{Ky} \ \geq \ \inp{\nabla \Psi(x)}{Ky}  \ =  \ K \omega \  > \ \orderednorm{x} + K + \NormApproxError \enspace .
\end{equation*}
However, by the norm approximation property,
\[
\Psi(x + Ky) \ \leq \ \orderednorm{x+Ky} + \NormApproxError \ \leq \ \orderednorm{x} + K + \NormApproxError \enspace ,
\]
which is a contradiction.
\end{proof}

Positivity of the gradients is a central property.
Indeed, many underlying combinatorial optimization problems, such as the shortest path problem, become much harder with negative prices.
The bounded dual norm of the gradients implies \emph{weak duality}:

\begin{corollary}[Weak duality]
\label{cor:inp_bounded_by_OPT}
Let $\Psi$ be a $( \NormApproxError, \eta)$-approximation of an ordered norm $\orderednorm{\cdot}$ on $\Rd$, 
let $X \subseteq \Rdpos$ and $x \in \Rdpos$.
Then, $\min_{s \in X} \inp{\nabla \Psi(x)}{s} \leq \min_{s \in X} \orderednorm{s}$.
\end{corollary}
\begin{proof}
For every $y \in \Rdpos$ with $\dualorderednorm{y} \leq 1$ and every $X \subseteq \Rdpos$,
\begin{equation}
	\label{eq:weak_duality}
	\min \left\{ \big. \inp{y}{s} : s \in X \right\} \
	\leq \ \min \left\{ \orderednorm{s} : s \in X \right\}  \enspace .
\end{equation}
Let $x \in \Rdpos$.
By \cref{lem:gradients_dual_norm_bounded_by_1},
$\dualorderednorm{\nabla \Psi(x)} \leq 1$.
Thus, the statement follows for $y=\nabla \Psi(x)$.
\end{proof}

\section{Warm-Up: Analyzing the Simple Algorithm in our Framework}
\label{sec:single_customer_continuous}
In this section, we present the simple algorithm sketched in \cref{sec:simplealganditslimitations} in detail
and analyze it with techniques that we will also apply to the general setting in \cref{sec:regret_based_ordered_norm_minimization_multiple_customers}.
This simple algorithm is well-known (see~\cite{agrawal2015fast} for example).

The algorithm uses $T$ iterations.
In each step $t = 1, \ldots, T$, we use a subroutine to determine a vector $y^{(t)}$, 
which can be understood as assigning a positive price to each resource. 
We then determine a solution $s^{(t)} \in X$ that minimizes $\langle y^{(t)}, s^{(t)} \rangle$.
In turn, $s^{(t)}$ is used by the subroutine to update the prices and obtain $y^{(t+1)}$. 
The final solution is given by the average of all solutions, $\bar{s} = \frac{1}{T} \sum_{t=1}^T s^{(t)}$.

In the following analysis, we want to compare $\orderednorm{\bar{s}}$ against $\OPT = \min_{x \in X} \orderednorm{x}$.
The algorithm and analysis will work with a smoothing parameter $\eta > 0$.
In this section, we restrict to bounded domains $X \subseteq [0,1]^d$ for simplicity, but will drop this assumption later in the generalized \cref{alg:general_ordered_norm_minimization_multiple_customers_continuous}.

\subsection{Simple Algorithm}\label{sec:simple_algorithm}

We use a piecewise linear function $S:[0,T] \to \Rdpos$ to keep track of the summed up solutions, 
where $S(0)=0^d$ and $S(t)-S(t-1)=s^{(t)}$ for $t \in \{1, \dots , T\}$.
For now, only the discrete values $S(t)$ for $t \in \left\{0,\ldots,T\right\}$ are relevant, but this will change later on.
Likewise, for now, working with Follow-the-Regularized-Leader (FTRL) price updates would be sufficient.
We decide for a different language that extends to the general setting in \cref{sec:regret_based_ordered_norm_minimization_multiple_customers} more easily.
The corresponding analysis based on the (standard) regret guarantee of FTRL is sketched in \cref{subsec:single_cust_ftrl_analysis}.
Here, we are using an $\bigl( \frac{\ln d}{\eta}, \eta \bigr)$-approximation $\Psi$ of the given ordered norm $\orderednorm{\cdot}$
to describe the prices $y^{(t)}$ in iteration $t$ as gradients of $\Psi$, 
i.e., $y^{(t)}=\nabla \Psi(S(t-1))$.
Such a norm approximation exists for all ordered norms as we will show in~\cref{sec:construction_norm_approximation}.

\begin{algorithm}[ht]
	\onehalfspacing
	\caption{Minimization of an ordered norm over $X$ with parameters $T \in \mathbb{N}$ and $\eta > 0$}
	\label{alg:ordered_norm_minimization_singe_customer_continuous}
	\begin{algorithmic}[1]
		\State $S(0) \gets 0^d$
		\For{$t = 1, \dots , T$}
		\State $s^{(t)} \ \gets \ \arg \min_{s \in X} \inp{\nabla \Psi( S(t-1) )}{s}$
		\Comment{Query oracle with prices $\nabla \Psi( S(t-1) )$.}
		\State Extend $S$ to the interval $[t-1, t]$ as a linear function from $S(t-1)$ to $S(t) \coloneqq S(t-1) + s^{(t)}$.
		\EndFor
		\State \Return $\bar{s} = \frac{S(T)}{T}$
	\end{algorithmic}
\end{algorithm}

\subsection{Performance Guarantee for Bounded Domains}

\begin{lemma}[Running time]
	Let $\orderednorm{\cdot}$ be an ordered norm on $\Rd$.
	For every choice of $T \in \mathbb{N}$ and $\eta > 0$,
	\cref{alg:ordered_norm_minimization_singe_customer_continuous} terminates after $T$ oracle calls and $T$ gradient evaluations. \hfill \qedsymbol
\end{lemma}

\begin{lemma}[Guarantee]
	\label{lem:guarantee_single_cust}
	Let $\orderednorm{\cdot}$ be an ordered norm on $\Rd$.
	Let $X \subseteq [0,1]^d$, $\eta > 0$, and $T \geq \frac{\ln d}{\eta^2}$. 
	For the result $\bar{s}$ of \cref{alg:ordered_norm_minimization_singe_customer_continuous}, we have
	\[ 
		\orderednorm{\bar{s}} \ 
		\leq \ \eta \ + \ \left( \frac{\exp(\eta)-1}{\eta} \right) \cdot \OPT \enspace .
	\]
	For $\eta \leq 1$, this implies $\orderednorm{\bar{s}} \leq \eta + (1+\eta) \cdot \OPT$.
\end{lemma}
\begin{proof}
	Note that by definition of $S$, for all $t \in \{ 0, \dots, T-1 \}$ and $w \in [0,1]$,
	$S(t+w) = S(t) + w \cdot s^{(t+1)}$.
	Therefore, due to bounded gradient increase of $\Psi$ with error $\eta$, and $\inftynorm{s} \leq 1$ for all $s \in X$, we have
	\begin{equation}
		\nabla \Psi(S(t+w)) \ \leq \ \exp \left( \eta \cdot w \cdot \inftynorm{s^{(t)}} \right) \cdot \nabla \Psi \left( S \left(t\right) \right) \ \leq \ \exp \left( \eta \cdot w \right) \cdot \nabla \Psi\left( S \left( t \right) \right) \enspace .
		\label{eq:proof_single_cust_exp_bound}
	\end{equation}
    Further, note that for $w \in (0,1)$, the derivative of $S$ is given by $S'(t+w)=s^{(t+1)}$ and $s^{(t+1)} \in X$ minimizes $\inp{\nabla \Psi (S(t))}{\cdot}$ over $X$.
	Thus, by \cref{cor:inp_bounded_by_OPT}, $ \inp{\nabla \Psi(S(t ))}{S'(t+w)} \leq \OPT$.
	Together, replacing $t+w$ by $z \in [0,T]$,
	\begin{equation}
		 \int_0^T \inp{\nabla \Psi(S(z))}{S'(z)} \, \text{d}z \
		 \stackrel{\text{\scriptsize \eqref{eq:proof_single_cust_exp_bound}}}{\leq} \ \int_0^T \exp \left( \eta \cdot \left(z - \lfloor z \rfloor \right) \right) \cdot \OPT \, \text{d}z  \
		 = \ T \cdot \OPT \cdot \left( \frac{\exp(\eta)-1}{\eta} \right) \enspace .
		 \label{eq:proof_single_cust_int_bound}
	\end{equation}
	
	By the error of our norm approximation, $\Psi(S(0)) = \Psi\left( 0^d \right) \leq \orderednorm{0^d} + \frac{\ln d}{\eta} = \frac{\ln d}{\eta}$.
	Combining this with \eqref{eq:proof_single_cust_int_bound} and the gradient theorem for line integrals on a (piecewise differentiable) curve,
	\begin{equation*}
		\Psi(S(T)) \ 
		= \ \Psi(S(0)) \ + \ \int_0^T \inp{\nabla \Psi(S(z))}{S'(z)} \, \text{d}z \
		\stackrel{\text{\scriptsize \eqref{eq:proof_single_cust_int_bound}}}{\leq} \ \frac{\ln d}{\eta} \ + \ T \cdot \OPT \cdot \left( \frac{\exp(\eta)-1}{\eta} \right) \enspace,
	\end{equation*}
	which implies, using $T \geq \frac{\ln d}{\eta ^2}$,
	\begin{equation*}
		\orderednorm{\bar{s}} \ = \ 
		\frac{\orderednorm{S(T)}}{T} \ 
		\leq \ \frac{\Psi(S(T))}{T} \ 
		\leq \ \eta \ + \ \left( \frac{\exp(\eta)-1}{\eta} \right) \cdot \OPT   \enspace .
	\end{equation*}
	For $\eta \leq 1$, we have $\frac{\exp(\eta)-1}{\eta} \leq 1 + \eta$, and thus $\orderednorm{\bar{s}} \leq \eta + (1+\eta) \cdot \OPT$.
\end{proof}

\cref{lem:guarantee_single_cust} provides a guarantee with additive and multiplicative error.
For $\OPT \geq 1$ and every $\varepsilon >0$, we can derive a multiplicative guarantee of $\orderednorm{\bar{s}} \leq \left( 1 + \varepsilon \right) \cdot \OPT$ after $\bigO{\frac{\ln d}{\varepsilon^2}}$ oracle calls and gradient evaluations.

Moreover, \cref{lem:guarantee_single_cust} assumes a bounded domain $X$. 
Alternatively, we can modify the algorithm by taking only a $\min\left\{1,\frac{1}{\inftynorm{s}}\right\}$ fraction of a solution $s$ returned by the oracle
(as suggested by \cite{garg2007faster} in a special case). This requires more oracle calls and price updates, but can be controlled
if $\OPT \leq 1$.

Together, we will need to assume $\OPT \approx 1$ to provide a multiplicative guarantee for $\orderednorm{\bar{s}}$ on unbounded domains $X$.
Combined with scaling and binary search, one can obtain an overall algorithm that proves \cref{thm:main} in the single-customer case $(n=1$).
To avoid repetitions, we do not elaborate the details here, but rather turn to the main difficulty:
handling multiple customers without calling all $n$ oracles in each iteration.

\section{Our General Algorithm for Multiple Customers}
\label{sec:regret_based_ordered_norm_minimization_multiple_customers}

We will now turn towards the setting with multiple customers that consume from the given set of resources.
This builds on \cref{sec:single_customer_continuous}, but several new ideas are needed. 
We are given a minimization oracle for each customer $c \in C$.
We may assume that no customer is trivial, meaning that $0^d \notin X_c$.
Trivial customers can be identified in advance (with a single oracle call) and their overall solution can be set to $0^d$.
We first present and analyze our core algorithm, which we later embed into an overall algorithm.

\subsection{Core Algorithm: Randomized Customer Selection and Adapted Step Sizes}

We introduce the randomized \cref{alg:general_ordered_norm_minimization_multiple_customers_continuous} for the multi-customer setting and unbounded domains.
We maintain a set of active customers $A \subseteq C$ over time.
Similar to the simple \cref{alg:ordered_norm_minimization_singe_customer_continuous} which runs for $T \in \mathbb{N}$ iterations,
we have a total work of $T$ that needs to be done for each customer.
We keep track of the remaining work to do for customers $c\in C$ by a vector $\rho\in\mathbb{R}^C_{\ge 0}$.
The amount $\xi$ of an oracle solution $s$ that is added depends on several properties,
including a well-chosen price bound,
and the total number of oracle calls for a customer can be fewer or more than $T$.
The price bound depends on a given guess $\GuessedOPT$ on $\OPT$.
Customers are sampled from the active set $A$ according to a distribution $\sigma$ 
that favors customers $c$ that have more remaining work $\rho_c$.
In fact, this distribution is proportional to the \emph{softmax} of $\rho$,
see \eqref{eq:probability_distribution}.
Once we have processed a total work of $T$ for a customer, we remove this customer from the active set.

\begin{algorithm}[ht]
	\onehalfspacing
	\caption{Core Algorithm}
	\label{alg:general_ordered_norm_minimization_multiple_customers_continuous}
	\vspace{.2cm}
	
	\hspace{.1cm} \textbf{Input:} First order oracle access to a norm approximation $\Psi$,\\
	\hspace*{1.47cm} linear minimization oracles for all $n$ customers $c \in C$,\\
	\hspace*{1.47cm} parameters $T \in \mathbb{N}$ , $\eta > 0$ , $\GuessedOPT > 0$ .
	\begin{algorithmic}[1]
		\State $A \ \gets \ C$ \Comment{Set of active customers}
		\State $\CustomerCurve(0) \ \gets \ (T, \dots, T) \in \mathbb{R}^C$ \Comment{Remaining work for each customer}
		\State $S(0) \ \gets \ 0^d$ \Comment{Collected solutions}
		\State $t \ \gets \ 0$ \Comment{Total collected work}
		\While{$A \neq \emptyset$}  \Comment{Equivalent to $t < nT$}
		\State $c \ \gets \ $ Sample a customer from $A$ with probabilities $\sigma_{\eta, A}(\CustomerCurve(t))$.\label{line:sample_c} 
		\Comment{See \eqref{eq:probability_distribution}}
		\State $s \ \gets \ \arg \min_{x \in X_c} \inp{\nabla \Psi( S(t) )}{x} $ \Comment{Query oracle with prices $\nabla \Psi( S(t) )$}
		\State $\Tilde{\xi} \ \gets \ \max \left\{0,\,  1 + \frac{1}{2 \eta} \cdot \ln \left( \frac{\GuessedOPT}{\inp{\nabla \Psi( S(t) )}{s}} \cdot \sigma_{\eta, A}(\CustomerCurve(t))_c \right) \right\}$
		\label{line:unnormalized_xi}
		\State $\xi \ \gets \min \left\{\Tilde{\xi}, \frac{\Tilde{\xi}}{\inftynorm{s}} \right\}$ \label{line:normalized_xi} \Comment{Garg--K\"onemann step size}
		\If{$\xi \geq  \CustomerCurve_c(t)$}
		\State $\xi \ \gets \  \CustomerCurve_c(t)$
		\State $A \ \gets \ A \setminus \{c\}$
		\EndIf
		\State Extend $S$ to the interval $[t, t+ \xi]$ as a linear function from $S(t)$ to $S(t + \xi) \coloneqq S(t) + \xi s$ .
		\State Extend $\CustomerCurve$ to the interval $[t, t+\xi]$ as a linear function from $\CustomerCurve(t)$ to $\CustomerCurve(t + \xi) \coloneqq \CustomerCurve(t) - \xi \mathbbm{1}_c$ .
		\State $t \ \gets \ t + \xi$
		\EndWhile
		\State \Return $\bar{s} = \frac{1}{T} S(nT)$
	\end{algorithmic}
\end{algorithm}

Note that Line 8 of \cref{alg:general_ordered_norm_minimization_multiple_customers_continuous} is well-defined, i.e., $\inp{\nabla \Psi(S(t))}{s} > 0$, because we excluded trivial customers and $\nabla \Psi > 0^d$ by \cref{lem:gradients_dual_norm_bounded_by_1}.

Depending on the guess $\GuessedOPT$ and the random choices in Line 6, 
\cref{alg:general_ordered_norm_minimization_multiple_customers_continuous}
may or may not terminate. 
Therefore, we will later stop the algorithm prematurely if it takes too many oracle calls 
and embed it into an overall algorithm that includes a binary search procedure
(to produce a good guess $\GuessedOPT\approx\OPT$) and a backup for unfavorable random choices.
As we will see, the algorithm is likely to terminate fast if $\GuessedOPT\ge\OPT$ 
and outputs a good solution if $\GuessedOPT$ is not much greater than $\OPT$.

\subsubsection{Sampling of Customers}
\label{subsubsec:customer_curve}

Consider the standard softmax function with smoothing parameter $\eta$, restricted to the set $A$ of active customers.
We denote it by $\sigma_{\eta, A}: \mathbb{R}^C_{\geq 0} \to \mathbb{R}^C_{> 0}$ with
\begin{equation}
	\sigma_{\eta, A}(z)_c \ \coloneqq \ \begin{cases}
		\ 0 & \text{for } c \notin A \\[.2cm]
		\ \frac{\exp(\eta z_c)}{\sum_{a \in A} \exp(\eta z_a)} & \text{for } c \in A
	\end{cases}
	\label{eq:probability_distribution}
\end{equation}
for $z \in \mathbb{R}^C_{\geq 0}$.
Let $A(t)$ denote the current set of active customers when a total work of $t$ has been processed.
We sample the next customer according to the probability distribution $\sigma_{\eta, A(t)}(\CustomerCurve(t))$.

Similary, we work with a variant of the \textsc{LogSumExp} function, denoted by $\LSE_{\eta, A} : \mathbb{R}^C_{\geq 0} \to \mathbb{R}_{\geq 0}$, with
\[
	\LSE_{\eta, A}(z) \ \coloneqq \ \frac{1}{\eta} \cdot \ln \ \sum_{c \in A}\exp \left( \eta z_c \right)
\]
for $z \in \mathbb{R}^C_{\geq 0}$.
It is easy to see that 
\begin{equation}
	\nabla \LSE_{\eta, A}(z) \ = \ \sigma_{\eta, A}(z) \enspace,
	\label{eq:gradient}
\end{equation}
which will play a crucial role in the analysis of our algorithm later.

\subsubsection{Solution Representation Through Piecewise Linear Curves}
\label{subsubsec:solution_curve}

We imagine that (rather than adding small pieces to our current solution) we walk from $0^d$ towards our final solution on a continuous piecewise linear curve $S: [0,nT] \rightarrow \mathbb{R}^d_{\ge 0}$.
$S(0) \coloneqq 0^d$ is our starting solution.
We then define the curve recursively through our algorithm in the following sense: 
When the algorithm adds an oracle solution $s$ with coefficient $\xi$ and the current total work is $t$,
we extend $S$ to the interval $[t,t+\xi]$ by appending a segment with the ``slope'' $s$ (that is, $S(t+\xi) = S(t) + \xi s$).
One can formally describe $S$ as 
\[ 
S(t) \ = \ \left(\int_{0}^t s_i(z) \, \text{d}z \right)_{i=1,\dots,d} \enspace ,
\]
where $s(z)$ denotes the oracle solution that was collected last before the total work exceeded $z$.
Note that only at termination, it is guaranteed that $\frac{1}{T}S(nT)$ is a feasible solution to the problem.
This does not hold necessarily for any other intermediate "average" vector $\frac{1}{t}S(nt)$, as it is often the case in previous round robin based approaches.

Similarly, we keep track of the remaining work for every customer.
We consider the function $\CustomerCurve:[0, nT] \rightarrow \mathbb{R}^C_{\geq 0}$ that stores the remaining work:
When the total work is $t$, we still need $\CustomerCurve(t)_c$ for customer $c$.
This is meant in the sense that we increase the work linearly for customer $c$ (and decrease $\CustomerCurve_c$ linearly with slope $-\mathbbm{1}_c$, where $\mathbbm{1}_c$ is the vector that contains a one in entry $c$ and zeros otherwise)
on the interval $[t,t+\xi]$, when we accept a solution for customer $c$ with amount $\xi$ at collection time $t$.
Note that $\CustomerCurve$ is a rectilinear curve in $\mathbb{R}^C_{\geq 0}$ going from $\CustomerCurve(0) = (T, \dots, T)$ to $\CustomerCurve(nT) = (0,\dots, 0)$.

\subsection{Analysis of \cref{alg:general_ordered_norm_minimization_multiple_customers_continuous}}

Our analysis of  \cref{alg:general_ordered_norm_minimization_multiple_customers_continuous}
consists of two main parts.
First, in \cref{thm:continuous_approx}, 
we make a deterministic statement about the approximation guarantee, which holds for all realizations of the algorithm,
assuming that the algorithm terminates.
The randomization only affects the running time, which we will analyze in \cref{section:running_time_of_core_algorithm}.

In the context of the bound on the approximation guarantee, 
it is helpful to view $\sigma$ purely analytically -- as the softmax of the curve $\rho$ -- rather than as a probability distribution.
We bound the total incurred cost by a path integral along $\rho$.
While different realizations of the algorithm may generate different monotonely decreasing curves $\rho$, 
the start and end points of $\rho$ are always the same, 
so we get a uniform upper bound on the path integral.

\subsubsection{Approximation Guarantee}

Rather than working with regret bound algorithms with discrete time steps, we utilize the gradient theorem for line integrals on a piecewise differentiable curve $S$, which states that for all $t \in [0, nT]$
\[
\Psi(S(t)) \ = \  \Psi(S(0)) +  \int_{0}^t \inp{\nabla \Psi(S(z))}{S'(z)} \, \text{d}z \enspace .
\]
In our specific case, $S'(z) = s(z)$ for all $z \in [0, nT]$ except for a finite number of points $z$ where $S(z)$ is not differentiable.
We may interpret the term $ \int_{0}^t \inp{\nabla \Psi(S(z))}{S'(z)} \, \text{d}z$ as the "cost incurred along the way".
Instead of the usual viewpoint, where we collect a solution $s$ with coefficient $\xi$ and incur a cost of $\inp{y}{\xi s}$ for the fixed price $y$ that was given as an input to the oracle, 
we pay for the solution $s$ under the evolving costs $\nabla \Psi(S(z))$ on an interval of length $\xi$.
To analyze these costs, the bounded increase of $\nabla \Psi$ is crucial.

\begin{theorem}[Guarantee]
    \label{thm:continuous_approx}
    Let $\orderednorm{\cdot}$ be an ordered norm on $\Rd$.
    Let $\NormApproxError \geq 0$, $\eta, \GuessedOPT > 0$, $T \in \mathbb{N}$, and let $\Psi$ be a $(\NormApproxError, \eta)$-approximation of $\orderednorm{\cdot}$.
    Assuming that \cref{alg:general_ordered_norm_minimization_multiple_customers_continuous} terminates, 
    the returned solution $\bar{s} = \frac{1}{T} S(nT)$ satisfies
    \begin{equation}
        \orderednorm{\bar{s}} \ \leq \ \exp(2\eta) \cdot \GuessedOPT \cdot \left( 1 + \frac{\ln n}{\eta T} \right) \ + \ \frac{\NormApproxError}{T} \enspace .
        \label{eq:approx_guarantee}
    \end{equation}
\end{theorem}

\begin{proof}
Set $z=nT$ in the following \cref{lemma:continuous_approx} and use $\orderednorm{\bar{s}}=\frac{1}{T}\orderednorm{S(nT)}$. 
\end{proof}

\begin{lemma}
    \label{lemma:continuous_approx}
    Let $\orderednorm{\cdot}$ be an ordered norm on $\Rd$.
    Let $\NormApproxError \geq 0$, $\eta, \GuessedOPT > 0$, $T \in \mathbb{N}$, and let $\Psi$ be a $(\NormApproxError, \eta)$-approximation of $\orderednorm{\cdot}$.
    For any $z\in[0,nT]$, we have in \cref{alg:general_ordered_norm_minimization_multiple_customers_continuous}: 
    \begin{equation*}
        \orderednorm{S(z)} \ \leq \ \exp(2\eta) \cdot \GuessedOPT \cdot \left( T + \frac{\ln n}{\eta} \right) \ + \ \NormApproxError \enspace .
    \end{equation*}
\end{lemma}

\begin{proof}
	We will drop the index $\eta$ from the notation of $\sigma_{\eta, A}$ and $\LSE_{\eta, A}$ for this proof.
	First, we use the aforementioned relation to the line integral:
	\begin{equation}
	    \orderednorm{S(z)} \
	    \leq \ \Psi(S(z)) \ = \ \Psi(S(0))  + \int_{0}^{z} \inp{\nabla \Psi(S(t))}{s(t)} \, \text{d}t \enspace.
	    \label{eq:x_integral_bound}
	\end{equation}
	Note that $\Psi(S(0)) = \Psi(0^d) \leq \NormApproxError$.
	Let $c(t) \in C$ be the customer corresponding to the solution $s(t)$ and $A(t)$ the corresponding set of active customers.
	We claim that for all $t \in [0, z)$, the incurred cost can be upper bounded by the corresponding entry of the softmax function according to which we sample, in the following sense:
	\begin{equation}
		\inp{\nabla \Psi(S(t))}{s(t)} \ \leq \ \exp(2\eta) \ \GuessedOPT \ \sigma_{A(t)}(\CustomerCurve(t))_{c(t)} \enspace.
		\label{eq:claimed_cost_bound_continuous}
	\end{equation}
	
	To prove \eqref{eq:claimed_cost_bound_continuous}, let us fix some value of $t$ in Line 6 of \cref{alg:general_ordered_norm_minimization_multiple_customers_continuous} for which the customer $c = c(t)$ is sampled according to $\sigma_{A(t)}(\CustomerCurve(t))$  and we attain a solution $s = s(t)$ that gets added with value $\xi > 0$.
	We will prove \eqref{eq:claimed_cost_bound_continuous} on the interval $[t, t+\xi)$.
	Then, \eqref{eq:claimed_cost_bound_continuous} must hold on the entire interval $[0, z)$ by the choice of $t$.
	In Line 8, $\Tilde{\xi}$ is chosen such that
	\[
	\Tilde{\xi} \ = \ 1 + \frac{1}{2 \eta} \cdot \ln \left( \frac{\GuessedOPT}{\inp{\nabla \Psi( S(t) )}{s}} \cdot \sigma_{A(t)}(\CustomerCurve(t))_c \right) \enspace ,
	\]
	which is equivalent to
	\begin{equation}
		\exp \bigl( 2 \eta \tilde{\xi} \bigr) \inp{\nabla \Psi(S(t))}{s} \ = \ \exp(2\eta) \ \GuessedOPT \ \sigma_{A(t)}(\CustomerCurve(t))_c \enspace .
		\label{eq:def_xi_tilde}
	\end{equation}
	Now let $\omega \in [0, \xi)$.
	For all such $\omega$, we have $s(t + \omega) = s$ and $c(t+\omega) = c$ and $A(t+\omega)=A(t)$. 
	Further, $\CustomerCurve(t + \omega)_{c} = \CustomerCurve(t)_c - \omega$.
	With this, we can derive
	\begin{equation}
		\sigma_{A(t+\omega)}(\CustomerCurve(t+\omega))_{c(t+\omega)} \ \geq \ \exp(-\eta \omega) \sigma_{A(t)}(\CustomerCurve(t))_{c} \enspace .
		\label{eq:sigma_lower_bound}
	\end{equation}
	Using the bounded gradient increase of $\Psi$ and the step size rule in Line 9, we can bound
	\begin{align}
		\inp{\nabla \Psi(S(t+\omega))}{s} \ 
		&= \ \inp{\nabla \Psi(S(t) + \omega s)}{s} \nonumber  \\[.2cm]
		&\leq \ \exp \bigl( \eta \omega \inftynorm{s} \bigr) \inp{\nabla \Psi(S(t)}{s} \nonumber \\[.2cm]
		&\leq \ \exp \bigl( \eta \tilde{\xi} \bigr) \inp{\nabla \Psi(S(t))}{s} \enspace . 
		\label{eq:gradient_upper_bound}
	\end{align}
	Together, we have
	\begin{align*}
		\inp{\nabla \Psi(S(t+\omega))}{s(t+\omega)} \ 
		& = \ \inp{\nabla \Psi(S(t+\omega))}{s} \\[.2cm]
		& \stackrel{\text{\scriptsize \eqref{eq:gradient_upper_bound}}}{\leq} \ \exp \bigl( \eta \tilde{\xi} \bigr) \inp{\nabla \Psi(S(t))}{s} \\[.2cm]
		& \stackrel{\text{\scriptsize \eqref{eq:def_xi_tilde}}}{=} \ \exp \bigl( -\eta \Tilde{\xi} \bigr) \exp(2\eta) \ \GuessedOPT \ \sigma_{A(t)}(\CustomerCurve(t))_c \\[.2cm]
		& \stackrel{\scriptstyle (\omega \leq \Tilde{\xi})}{\leq} \exp(-\eta \omega) \exp(2\eta) \ \GuessedOPT \ \sigma_{A(t)}(\CustomerCurve(t))_c \\[.2cm]
		&\stackrel{\text{\scriptsize \eqref{eq:sigma_lower_bound}}}{\leq} \ \exp(2\eta) \ \GuessedOPT \ \sigma_{A(t+\omega)}(\CustomerCurve(t+\omega))_{c(t+\omega)} \enspace ,
	\end{align*}
	which proves our claim \eqref{eq:claimed_cost_bound_continuous}.
	
	Let $a_0 =0 < a_1 < \dots < a_n = nT$ be the values of $t$ at which the size of $A$ changes. This happens only in Line 12 and we have $|A(a_i)| = n - i$.
	Now the approximation guarantee follows easily by comparing to an appropriate line integral of the softmax function:
	{
		\allowdisplaybreaks
		\begin{align}
		    \int_0^{z} \inp{\nabla \Psi(S(t))}{s(t)} \, \text{d}t \ \nonumber
		    & \stackrel{\text{\scriptsize \eqref{eq:claimed_cost_bound_continuous}}}{\leq} \ \exp(2 \eta) \,\GuessedOPT \int_0^{z} \sigma_{A(t)}(\CustomerCurve(t))_{c(t)} \, \text{d}t \nonumber \\[.3cm]
		    & \leq \ \exp(2 \eta) \,\GuessedOPT \int_0^{nT} \sigma_{A(t)}(\CustomerCurve(t))_{c(t)} \, \text{d}t \nonumber \\[.3cm]
		    & = \ \exp(2 \eta) \,\GuessedOPT \int_0^{nT}  \inp{\sigma_{A(t)}(\CustomerCurve(t))}{\mathbbm{1}_{c(t)}} \, \text{d}t \nonumber \\[.3cm]
		    & \stackrel{\text{\scriptsize \eqref{eq:gradient}}}{=} \ \exp(2 \eta) \,\GuessedOPT \int_0^{nT} - \inp{\nabla \LSE_{A(t)}(\CustomerCurve(t))}{\CustomerCurve'(t)} \, \text{d}t \nonumber \\[.3cm]
		    & = \ \ \exp(2 \eta) \,\GuessedOPT \,\sum_{i=0}^{n-1} \ \left( \int_{a_i}^{a_{i+1}} - \inp{\nabla \LSE_{A(t)}(\CustomerCurve(t))}{\CustomerCurve'(t)} \, \text{d}t \right) \nonumber \\[.3cm]
		    & = \ \exp(2 \eta) \,\GuessedOPT \,\sum_{i=0}^{n-1} \left( \LSE_{A(a_i)}(\CustomerCurve(a_i)) - \LSE_{A(a_i)}(\CustomerCurve(a_{i+1})) \right) \nonumber \\[.3cm]
		    & \leq \ \exp(2 \eta) \,\GuessedOPT \ \LSE_{A(a_0)}(\CustomerCurve(a_0)) \nonumber \\[.3cm]
		    & = \ \exp(2 \eta) \,\GuessedOPT \ \LSE_C \left( T \cdot 1^n \right) \nonumber \\[.3cm]
		    & = \ \exp(2 \eta) \,\GuessedOPT \left( T + \frac{\ln n}{\eta}\right) \enspace .
		    \label{eq:integral_bound}
		\end{align}
	}
	The last inequality follows from the fact that the $\LSE$ function increases with the dimension, i.e., $\LSE_{A(a_i)}(\CustomerCurve(a_{i+1})) \geq \LSE_{A(a_{i+1})}(\CustomerCurve(a_{i+1}))$.
	Finally, inserting \eqref{eq:integral_bound} into \eqref{eq:x_integral_bound} together with $\Psi(S(0)) \leq \NormApproxError$ proves the claimed bound.
\end{proof}

We stress once again that any order of processing customers yields the claimed approximation guarantee.
However, an unfavorable order may lead frequently to tiny steps or even rejection ($\tilde{\xi}=0$) of the oracle solution
even if $\GuessedOPT\ge\OPT$.
Our non-uniform sampling strategy avoids this issue and ensures the desired bound on the number of oracle calls in expectation, 
as we will show next.
In prior work on the $\ell_{\infty}$ norm, sufficient progress could be guaranteed by round robin schemes \cite{khandekar2004lagrangian, young2001sequential}; however, this does not work for general ordered norms.

\subsubsection{Running Time} \label{section:running_time_of_core_algorithm}

Since each iteration of \cref{alg:general_ordered_norm_minimization_multiple_customers_continuous}
contains one gradient evaluation and one oracle call, we want to bound the number of iterations.
We will do this under the assumption $\GuessedOPT \geq \OPT$. 
In the end, we will stop the algorithm prematurely if it takes too many iterations, 
and then conclude $\GuessedOPT < \OPT$ with high probability. 
We will remove the remaining minor failure probability in \cref{section:backup_for_unlucky_random_choices}.

The core of the proof of the bound on the number of iterations consists of showing that, assuming $\GuessedOPT \geq \OPT$,
the expected progress in most iterations, regardless of what happened before, is at least 1.
There can be exceptions in iterations when $\xi<\tilde{\xi}$, but we will see soon that there are not too many of these. 
First, we consider the conditional expectation of $\tilde{\xi}$, conditioned on the past and capped at $1 + \frac{1}{2\eta}$:
More precisely,
for each iteration $i$ of \cref{alg:general_ordered_norm_minimization_multiple_customers_continuous}, let
	\[
		Y_i \ \coloneqq \ \min\left\{ 1 + \frac{1}{2\eta} \ , \ \tilde{\xi}_i \right\} \enspace,
	\] 
where $\tilde{\xi}_i$ denotes the value of $\tilde{\xi}$ in Line 8 of \cref{alg:general_ordered_norm_minimization_multiple_customers_continuous} in the $i$-th iteration of the while-loop.
To be able to reasonably discuss expected values of the $Y_i$ over different realizations of the algorithm with different numbers of iterations, 
we define $Y_i \coloneqq 1$ if $i$ exceeds the number of iterations.

\begin{lemma}\label{lemma:enoughprogress}
If $\GuessedOPT \geq \OPT$, then $\Exp \left[ Y_i \, | \, \text{iterations } 1,\ldots, i-1 \right] \ge 1$ for all $i\in\mathbb{N}$, regardless of what happened in iterations $1,\ldots,i-1$ of \cref{alg:general_ordered_norm_minimization_multiple_customers_continuous}.
\end{lemma}

\begin{proof}
	Again, we will drop the index $\eta$ from the notation of $\sigma_{\eta, A}$ for this proof.
	For every iteration $i$ of \cref{alg:general_ordered_norm_minimization_multiple_customers_continuous}, we have for $A$, $c$ and $s$ as defined in that iteration:  

    \begin{align*}
		Y_i \ &\geq \ \min \left\{ 1 + \frac{1}{2 \eta} \ , \  1 + \frac{1}{2\eta} \cdot \ln \left( \frac{\GuessedOPT \cdot \sigma_A(\CustomerCurve(t))_c}{\inp{\nabla \Psi(S(t))}{s}}\right) \right\} \\[.3cm]
		&= \ 1 \ + \ \frac{1}{2 \eta} \cdot \min \left\{ 1 \ , \ \ln \left( \frac{\GuessedOPT \cdot \sigma_A(\CustomerCurve(t))_c}{\inp{\nabla \Psi(S(t))}{s}}\right) \right\} \\[.3cm]
		&\geq \ 1 \ + \ \frac{1}{2 \eta} \left( 1 - \frac{\inp{\nabla \Psi(S(t))}{s}}{\GuessedOPT \cdot \sigma_A(\CustomerCurve(t))_c} \right) \enspace.
	\end{align*}
    The last inequality follows from the property of the natural logarithm $\ln (z) \geq 1- \frac{1}{z}$ for all $z > 0$.
	Fix the first $i-1$ iterations of \cref{alg:general_ordered_norm_minimization_multiple_customers_continuous}. 
	Note that this corresponds to fixing a customer sequence.
	This uniquely defines the values of $S$, $\CustomerCurve$, $A$ and $\sigma_A$ at the end of iteration $i-1$.
	If $A = \emptyset$, we have $Y_i=1$ by definition. 
	
	Otherwise, if $A \neq \emptyset$, there will be another iteration $i$, using these values. 
	We will randomly choose a customer $c$ from $A$, which defines a solution $s_c$. 
	The above calculation now implies that
	\begin{align}
		\Exp \left[\big. Y_i \ | \ \text{iterations } 1,\ldots,i-1 \ , \ A \neq \emptyset \big.\right] \
		&\geq \ \sum_{c \in A} \sigma_A(\CustomerCurve(t))_c \left( 1 +  \frac{1}{2 \eta} \left( 1 - \frac{\inp{\nabla \Psi(S(t))}{s_c}}{\GuessedOPT \cdot \sigma_A(\CustomerCurve(t))_c} \right) \right)  \notag \\[.3cm]
		&= \ 1 \ + \ \frac{1}{2\eta} \ - \ \frac{1}{2\eta} \cdot \frac{1}{\GuessedOPT} \cdot \Bigg(\underbrace{ \sum_{c \in A} \inp{\nabla \Psi(S(t))}{s_c} }_{\leq \ \OPT, \text{ by \cref{cor:inp_bounded_by_OPT}}} \Bigg) \notag \\[.3cm]
		&\geq \ 1 \ + \ \frac{1}{2\eta} \ - \ \frac{1}{2\eta} \cdot \frac{\OPT}{\GuessedOPT} \notag \\[.3cm]
		&\geq \ 1 \enspace,   \label{eq:expectation_at_least_one}
	\end{align}
	where we used $\GuessedOPT \geq \OPT$ in the last step. 
\end{proof}

To prove an upper bound on the running time, we also need to handle iterations with $\xi<\tilde{\xi}$.
Let
	\[
		\Omega \ \coloneqq \ nT +  d \left( \exp(2\eta) \ \GuessedOPT \ \left( T + \frac{\ln n}{\eta} \right) + \NormApproxError \right) + \frac{n}{2\eta} + n \enspace.
	\]
Eventually, we will combine \cref{lemma:enoughprogress} with the following.

\begin{lemma} \label{lemma:bound_sum_of_Y_i}
For every iteration $j$ of \cref{alg:general_ordered_norm_minimization_multiple_customers_continuous},
	\[
		\sum_{i=1}^{j} Y_i \ \leq \ \Omega  \enspace .
	\]
\end{lemma}

\begin{proof}	
	For each iteration $i$ of the while-loop, let $\xi_i$ and $t_i$ and $A_i$ denote the values of $\xi$ and $t$ and $A$ at the end of this iteration.
	We consider how the following potential function changes in one iteration:
	\[ 
		\Phi_j \ \coloneqq \ \sum_{i=1}^j \xi_i \ + \ \onenorm{S(t_j)} \ + \ \left(1+\frac{1}{2\eta}\right) (n-|A_j|) \enspace .
	\]
	Consider iteration $j$. If $\xi_j<\tilde{\xi}_j$, then 
	either (due to Line 9) $\onenorm{S(t_j)} \geq \tilde{\xi}_j + \onenorm{S(t_{j-1})}$,
	or (due to Lines 11 and 12) the customer that is sampled in iteration $j$ becomes inactive (or both). 
	In every case, 
	\[
	\Phi_j-\Phi_{j-1} \ \ge \ \min \left\{ 1+\frac{1}{2\eta} \ , \ \tilde{\xi}_i \right\} \ = \ Y_i \enspace.
	\]
	Therefore, using $\Phi_0=0$, we have for all iterations $j$ of the while-loop,
	\begin{equation*}
		\sum_{i=1}^j Y_i \ \leq \ \Phi_j  \enspace.
	\end{equation*}
	Moreover, for all iterations $j$ of the while-loop,
	\begin{align*}
		\Phi_j \ 
		&= \ t_j + \onenorm{S(t_j)} + \left(1+\frac{1}{2\eta}\right) (n-|A_j|) \\
		&\leq \ nT + d \orderednorm{S(t_j))} +n+\frac{n}{2\eta} \\
		&\leq \ nT + d \left( \exp(2\eta) \GuessedOPT  \left( T + \frac{\ln n}{\eta} \right) + \NormApproxError  \right) +n+\frac{n}{2\eta} \enspace,
	\end{align*}
	where the first inequality follows from $\onenorm{x}  \leq d \cdot \orderednorm{x}$ (\cref{prop:l1_bound}),
	and the last inequality follows from \cref{lemma:continuous_approx} for $z=t_j$. 
\end{proof}

\cref{lemma:enoughprogress} and \cref{lemma:bound_sum_of_Y_i} can be combined to bound the running time.
It is very unlikely that the algorithm takes more than $2\Omega$ iterations (assuming $\GuessedOPT \geq \OPT$; cf.\ \cref{lemma:enoughprogress}):

\begin{lemma}
	\label{lemma:continuous_runtime}
	Let $\orderednorm{\cdot}$ be an ordered norm on $\Rd$.
	Let $\NormApproxError \geq 0$, $\eta > 0$, $T \in \mathbb{N}$, $\Gamma>0$, 
	and let $\Psi$ be a $(\NormApproxError, \eta)$-approximation of $\orderednorm{\cdot}$.
	The number of iterations $i$ of \cref{alg:general_ordered_norm_minimization_multiple_customers_continuous} 
	in which $\Exp \left[\big. Y_i \ | \ \text{iterations } 1,\ldots,i-1 \right] \ge 1$ exceeds $2\Omega$
	with probability at most $\exp \left(- \Omega \cdot \min \left\{ \eta^2,\frac{1}{4} \right\} \right)$.
\end{lemma}

\begin{proof}
	Note that only the choice of a customer per iteration is non-deterministic.
	Consider the sequence of random variables $Y_1, Y_2, \dots$ defined as above, and let 
	$I=\{j_1,\ldots,j_{|I|}\}$ denote the set of the first $2\Omega$ iterations $i$ 
	in which the conditional expected value of $Y_i$ (conditioned on all previous iterations) is at least 1.
	Restricting the conditioning to the prior outcomes of $Y_i$ yields, for $i\in I$ and all $y_1,\ldots,y_{i-1}$,
	\[
		\Exp \left[\big. Y_i \ | \ Y_{i-1} = y_{i-1}, \dots, Y_1 = y_1 \big.\right] \ \geq \ 1 \enspace .
	\]
	This implies that $M_l := \sum_{i=1}^l \left( Y_{j_i} -1 \right)$ ($l\in\mathbb{N}$) is a submartingale, 
	with (absolute) differences bounded by $\max \left\{ \frac{1}{2\eta},1 \right\}$ since $Y_i\in \left[0, 1+\frac{1}{2 \eta} \right] $.
	By the Azuma--Hoeffding inequality \cite{azuma1967weighted}, and using $|I| = 2\Omega$, we get that
	\allowdisplaybreaks
	\begin{align}
		\Pr\left[ \sum_{i\in I} Y_i \leq \Omega \right] \
		&= \ \Pr\left[\big. M_{|I|} \leq \Omega-|I| \big.\right]  \nonumber \\
		&\leq \ \exp \left( - \frac{ (\Omega-|I|)^2}{ 2 |I| \cdot \left( \max \left\{ \frac{1}{2\eta},1 \right\} \right)^2 } \right) \nonumber \\[.3cm]
		&= \ \exp \left( -\Omega \cdot \min \left\{ \eta^2,\frac{1}{4} \right\} \right).
		\label{eq:azuma_hoeffding} 
	\end{align}
	By \cref{lemma:bound_sum_of_Y_i}, the algorithm terminates 
	before the sum of the $Y_i $ exceeds $\Omega$.
\end{proof}

Combining this with \cref{lemma:enoughprogress} yields an upper bound of $2\Omega$ on the number of iterations
with high probability, assuming $\GuessedOPT \geq \OPT$.

\begin{theorem}[Running time of \cref{alg:general_ordered_norm_minimization_multiple_customers_continuous}]
	\label{thm:continuous_runtime}
	Let $\orderednorm{\cdot}$ be an ordered norm on $\Rd$.
	Let $\NormApproxError \geq 0$, $\eta > 0$, $T \in \mathbb{N}$, and let $\Psi$ be a $(\NormApproxError, \eta)$-approximation of $\orderednorm{\cdot}$.
	Let $\GuessedOPT \geq \OPT$.
	Then the number of iterations of \cref{alg:general_ordered_norm_minimization_multiple_customers_continuous} 
	exceeds $2\Omega$
	with probability at most $\exp \left(- \Omega \cdot \min \left\{ \eta^2,\frac{1}{4} \right\} \right)$.
\end{theorem}

\begin{proof}
By \cref{lemma:enoughprogress}, we have $\Exp \left[\big. Y_i \ | \ \text{iterations } 1,\ldots,i-1 \right] \ge 1$ in every iteration, so this follows
directly from \cref{lemma:continuous_runtime}.
\end{proof}

For later use, we also consider a slower deterministic variant of \cref{alg:general_ordered_norm_minimization_multiple_customers_continuous}:
	
\begin{theorem}[Running time of deterministic variant]
\label{thm:continuous_runtime_deterministic}
Consider the following variant of \cref{alg:general_ordered_norm_minimization_multiple_customers_continuous}: 
in each iteration, we call all $n$ oracles and take the first active customer for which $\tilde{\xi}\ge 1$
(and stop if there is no such customer). 
This variant then terminates after at most $\Omega$ iterations.
\end{theorem}	

\begin{proof}
This follows directly from \cref{lemma:bound_sum_of_Y_i} because $\tilde{\xi}_i\ge 1$ implies $Y_i\ge 1$.
\end{proof}

Of course, this requires $n$ oracle calls per iteration, which we want to avoid.

\subsection{Overall Algorithm} 

Our overall algorithm contains two additional components besides the core algorithm.
First, we design a variant of \cref{alg:general_ordered_norm_minimization_multiple_customers_continuous}
that finds a solution $\bar s$ with $\orderednorm{\bar s} \le (1+3\eta)\GuessedOPT$
or decides that $\GuessedOPT < \OPT$, and is guaranteed to terminate.
Then we will use this in a binary search framework (to obtain $\GuessedOPT \approx \OPT$).

\subsubsection{Backup for Unlucky Random Choices}
\label{section:backup_for_unlucky_random_choices}

If \cref{alg:general_ordered_norm_minimization_multiple_customers_continuous} takes more than
$2\Omega$ iterations, it is likely that $\GuessedOPT<\OPT$, but it is not completely sure.
Although the failure probability is very low, it is worthwhile to eliminate it completely.
This is achieved by \cref{alg:main_algorithm_with_guaranteed_termination}.
It consists of three parts. 
First, it runs \cref{alg:general_ordered_norm_minimization_multiple_customers_continuous} for $3\Omega$ iterations.
If it does not terminate, this means with high probability that
$\Exp \left[\big. Y_i \ | \ \text{iterations } 1,\ldots,i-1 \right] < 1$ in many iterations (and hence $\GuessedOPT < \OPT$).
Then, in the second part, we try to identify one of these iterations in order to prove $\GuessedOPT < \OPT$.
If we do not find one in $\frac{\Omega}{n}$ random attempts, we must be very unlucky.
In this case, we use the slow deterministic variant as a backup; this third part is executed only with very low probability
(or if $n\le 4$, in which case it is fast enough).

\begin{algorithm}[ht]
	\onehalfspacing
	\caption{Main Algorithm with Guaranteed Termination}
	\label{alg:main_algorithm_with_guaranteed_termination}
	\vspace{.2cm}
	
	\hspace{.1cm} \textbf{Input:} First order oracle access to a $(\delta,\eta)$-approximation $\Psi$,\\
	\hspace*{1.47cm} linear minimization oracles for all $n$ customers $c \in C$,\\
	\hspace*{1.47cm} parameters $T \ge \frac{4}{\eta^2}\ln(nd)$, $0 < \eta \le \frac{1}{5}$, $\GuessedOPT > 0$.
	\begin{algorithmic}[1]
		\State $\Omega	 \ \gets \ nT +  d \left( \exp(2\eta) \GuessedOPT \left( T + \frac{\ln n}{\eta} \right) + \NormApproxError \right) + \frac{n}{2\eta}+n$.
		\If{$n\ge 5$}
		
\Comment First try \cref{alg:general_ordered_norm_minimization_multiple_customers_continuous}.			
		\State Run \cref{alg:general_ordered_norm_minimization_multiple_customers_continuous} for up to $3\Omega$ iterations and store  
		$\rho(t_{i-1})$ and $\nabla \Psi( S(t_{i-1}) )$ for $i=1,\ldots,3\Omega$.
		\If{the algorithm terminates with $A=\emptyset$ and a feasible solution $\bar s$} 
		\State{\textbf{stop} and output $\bar s$.}
		\EndIf

\Comment{Try to find an iteration that proves $\GuessedOPT < \OPT$.}	
		\For{$j=1,\ldots,\frac{\Omega}{n}$}
		\State{pick $i\in\{1,\ldots,3\Omega\}$ uniformly at random}
		\If{$\Exp\left[\big. Y_i \ | \ \text{iterations } 1 \text{ to } i-1 \big.\right] < 1$} 
		\State{\textbf{stop} and conclude that $\GuessedOPT < \OPT$.}
		\EndIf
		\EndFor
		\EndIf
		
\Comment Backup run that is slow but will always work.		
		\State Run \cref{alg:general_ordered_norm_minimization_multiple_customers_continuous}, 
		 but instead of sampling $c$ in Line 6, take an active customer 
		 for which $\tilde{\xi} \ge 1$. 
		\If{there is no such $c$ in some iteration}
		\State \textbf{stop} and conclude that $\GuessedOPT < \OPT$.
		\EndIf
		\State{Otherwise the algorithm will terminate with a feasible solution $\bar s$; output $\bar s$.}
	\end{algorithmic}
\end{algorithm}

\begin{corollary}\label{cor:guaranteed_behaviour}
If \cref{alg:main_algorithm_with_guaranteed_termination} stops prematurely in Line 10 or 16, then $\GuessedOPT < \OPT$. 
Otherwise, the result $\bar s$ obeys the guarantee \eqref{eq:approx_guarantee}.
The algorithm makes at most $(4+n)\Omega$ oracle calls, and at most $4\Omega$ gradient evaluations of $\Psi$.
It makes more than $4\Omega$ oracle calls with probability at most $2(nd)^{-20}$. 
\end{corollary}

\begin{proof}
\cref{thm:continuous_runtime_deterministic} guarantees that
the slow variant of \cref{alg:general_ordered_norm_minimization_multiple_customers_continuous}
that we may call in Line 14 terminates after $\Omega$ iterations.
Each such iteration contains one gradient evaluation and up to $n$ oracle calls.
This already implies the claim for $n\le 4$.

Line 3 requires storing $\bigO{3 \Omega (n+d)}$ numbers: 
in each iteration $i$ we store the ``remaining work vector'' $\rho(t_{i-1})\in\mathbb{R}^C_{\ge 0}$ and
the price vector $\nabla \Psi( S(t_{i-1}) ) \in \Rdpos$, 
where $t_{i-1}$ again denotes the value of $t$ at the beginning of iteration $i$.
Line 9 requires calling the oracle with the stored price vector $\nabla \Psi( S(t_{i-1}) )$ for each active customer, computing 
$Y_i=\min\{1+\frac{1}{2\eta},\tilde{\xi}_i\}$ for this customer, and taking the weighted average.
Hence the maximum number of oracle calls is $3\Omega$ is Line 3, $\Omega$ in Line 9, and $n\Omega$ in Line 14.
There is only one gradient evaluation in each of the at most $3\Omega$ iterations in Line 3 and in each of the at most $\Omega$ iterations in Line 14. 

If the algorithm stops prematurely in Line 10 or 16, then
$\Exp \left[\big. Y_i \ | \ \text{iterations } 1 \text{ to } i-1 \big.\right] < 1$, 
and hence $\OPT>\Gamma$ by \cref{lemma:enoughprogress}.
Otherwise, the call to \cref{alg:general_ordered_norm_minimization_multiple_customers_continuous} 
in Line 3 or 14 succeeds, and \cref{thm:continuous_approx} guarantees \eqref{eq:approx_guarantee}.

It remains to bound the expected number of oracle calls. 
If the algorithm terminates in Line 5, we are done.
Otherwise, we distinguish two cases.

\noindent \emph{Case 1:} 
When running \cref{alg:general_ordered_norm_minimization_multiple_customers_continuous} in Line 3,
we have $\Exp \left[\big. Y_i \ | \ \text{iterations } 1 \text{ to } i-1 \big.\right] \ge 1$
in more than $2\Omega$ iterations $i$.
By \cref{lemma:continuous_runtime}, the probability that this happens is at most
$\exp \left( - \eta^2 \Omega \right) \le (nd)^{-4n} \le (nd)^{-20}$,
where we used $\Omega\ge nT\ge \frac{4n}{\eta^2}\ln(nd)$.

\noindent \emph{Case 2:} 
When running \cref{alg:general_ordered_norm_minimization_multiple_customers_continuous} in Line 3,
we have $\Exp \left[\big. Y_i \ | \ \text{iterations } 1 \text{ to } i-1 \big.\right] < 1$
in at least $\Omega$ iterations $i$.
Note that \cref{alg:main_algorithm_with_guaranteed_termination} will stop in Line 10 once we process such an iteration.
The probability that $\Exp \left[\big. Y_i \ | \ \text{iterations } 1 \text{ to } i-1 \big.\right] \ge 1$ in Line 9 
for all $\frac{\Omega}{n}$ iterations 
(and hence the probability that we need more than $4\Omega$ oracle calls in total in this case) is at most 
\[
\left(\frac{2}{3}\right)^{\frac{\Omega}{n}} \ = \ \exp \left( - \frac{\Omega}{n} \ln \frac{3}{2} \right) \ \le \ (nd)^{-20} 
\]
using $\Omega\ge nT\ge 100n \ln(nd)$.

Taking the union bound of the two cases concludes the proof.
\end{proof}

If we set $T$ appropriately, we get:

\begin{corollary}\label{cor:guaranteed_behaviour_nice}
Let $\Psi$ be an $\left( \frac{\ln d}{\eta}, \eta \right)$-approximation of $\orderednorm{\cdot}$.
Then there is an algorithm that, given any instance, any $\eta$ with $0<\eta\leq\frac{1}{5}$, and any $\GuessedOPT>0$,
decides that $\OPT>\Gamma$ or computes a solution $\bar s\in X$ with $\orderednorm{\bar s} \le (1+3\eta)\GuessedOPT$,
and makes at most $4\Omega$ gradient evaluations of $\Psi$, at most $(4+n)\Omega$ oracle calls,
and with probability at least $1-2(nd)^{-20}$ at most $4\Omega$ oracle calls,
where $\Omega\in\bigO{\frac{n+d}{\eta^2}(\log (n+d)}$.
\end{corollary}

\begin{proof}
We may assume $\GuessedOPT=1$ because we can simply scale the instance by $\frac{1}{\GuessedOPT}$.
We apply \cref{cor:guaranteed_behaviour} to $T=\frac{4}{\eta^2}\ln (nd)$. 
Then \eqref{eq:approx_guarantee} implies
\[
\orderednorm{\bar s} \ \le \ \exp(2\eta)\cdot \left(1+\frac{\eta}{4}\right) 
\ \le \ 1+3\eta.
\]
Moreover, \cref{cor:guaranteed_behaviour} implies the stated bounds for
\[
\Omega \ = \ nT +  d \left( \exp(2\eta) \ \GuessedOPT \left( T + \frac{\ln n}{\eta} \right) + \frac{\ln d}{\eta} \right) + \frac{n}{2\eta}+n
\ \in \ \bigO{\frac{n+d}{\eta^2} \log (n+d)} \enspace . \qedhere 
\]
\end{proof}

\subsubsection{Binary Search}

Finally, we embed \cref{cor:guaranteed_behaviour_nice} in a binary search framework and obtain our main theorem.

\begin{theorem}[Main theorem]
    \label{thm:main_theorem_precise}
    Let $\orderednorm{\cdot}$ be an ordered norm on $\Rd$ with an $\left( \ln d, 1\right)$-approximation $\Psi$.
    Then there is an algorithm that, for any given instance and any $\varepsilon \in (0,1]$,
    computes a solution $x \in X$ with $\orderednorm{x} \leq (1+\varepsilon) \cdot \OPT$, 
    and makes at most $\Omega^*$ gradient evaluations of $\Psi$, at most $n\Omega^*$ oracle calls, 
    and more than $\Omega^*$ oracle calls with probability at most $(nd)^{-18}$, where
    \[
     \Omega^*\ \in \ \bigO{ (n+d) \log (n+d) \cdot \left( \varepsilon^{-2} + \log \log d \right) } \enspace .
    \]
\end{theorem}

\begin{proof}
    First we remark that for any $\eta>0$, we have an $\left( \frac{\ln d }{\eta}, \eta\right)$-approximation
    $x\mapsto\frac{1}{\eta}\Psi (\eta x)$ of $\orderednorm{\cdot}$. We will use this for various values of $\eta$.

    Our goal is to use \cref{cor:guaranteed_behaviour_nice} 
    with some sufficiently small $\eta$ and $\GuessedOPT \approx \exp(\eta) \cdot \OPT$.
    To do this, we need to approximate $\OPT$, which is possible with a two-stage binary search.
    
    First, as a preparation step, we call the oracle for every customer with uniform prices $1^d$ in order to get a solution $v \in X$ with minimal $\ell_1$ norm.
    By \cref{prop:l1_bound}, we have
    \[
    	\OPT \ \leq \ \orderednorm{v} \ \leq \ \inftynorm{v} \leq \ \onenorm{v} \ \leq \ d \cdot \OPT \enspace .
  	\] 
    Thus, we have an initial guess $\orderednorm{v}$ of $\OPT$ that is off by at most a factor $d$
    (and we are already done if $d=1$).

    In the first stage of our binary search, we compute a constant-factor approximation of $\OPT$.
    We create $\lceil \log_2 d \rceil$ scaled instances $X^{(1)}, X^{(2)}, \dots, X^{(\lceil \log_2 d \rceil)}$ with $X^{(i)} \coloneqq \frac{2^i}{\orderednorm{v}} \cdot X$.
    Let us denote with $\OPT^{(i)}$ the value of $\OPT$ on instance $X^{(i)}$ for $i \in \left\{1, \ldots, \lceil \log_2 d \rceil \right\}$.
    Note that $\OPT^{(i)} = 2 \cdot \OPT^{(i-1)}$.
    This first stage of binary search will identify an index $i$ such that $1 \leq \OPT^{(i)} \leq 4$.
    It exists by our initial $d$-approximation and the chosen scaling.
    We will determine the largest index $i$ for which the algorithm of \cref{cor:guaranteed_behaviour_nice}
    applied to $X^{(i)}$ and $\GuessedOPT^{(i)}=2$ and $\eta=\frac{1}{5}$
    terminates with a solution $\bar s$.

	\cref{cor:guaranteed_behaviour_nice} guarantees to compute either 
  	$\bar{s} \in X$ with $\orderednorm{\bar{s}} \leq 4$ (so $\OPT^{(i)} \le 4$; in this case we will not consider smaller indices than $i$ in the following)
  	or decides that $\OPT^{(i)} > 2$ for the current index $i$, implying that $\OPT^{(i-1)} > 1$ (in this case we will only consider smaller indices than $i$ in the following).
    In total, we conclude that this first stage of binary search finds an index $i$ with $\OPT^{(i)} \in [1, 4]$.
It needs $\bigO{ (n+d)\log (n+d) \log \log d }$ gradient evaluations,
and with high probability not more oracle calls (we will bound this probability at the end of the proof).
 
    For the remainder of the proof, we assume that we have an instance with $\OPT \in [1,4]$.
    We aim to construct a sequence of intervals $I_j = [a_j, b_j] \subseteq [1,4]$ such that $\OPT \geq a_j$ for all $j$ 
    and we always have a solution $\bar{s}$ with norm $\orderednorm{\bar{s}} \leq b_j$ from the previous iteration,
    and the size of the intervals decreases exponentially.
    More precisely, the intervals will satisfy 
    \[
    	b_j - a_j \ \leq \ \frac{2}{3} \left( b_{j-1} - a_{j-1} \right) \enspace .
   	\]
    Once $b_j - a_j \leq \varepsilon$, we terminate the binary search and return the best solution $x \in X$ found so far.
    By construction of the binary search, 
    we can bound its objective value by $\orderednorm{x} \leq b_j \leq a_j+\varepsilon \leq (1+\varepsilon) \OPT$.

    We start the second stage of the binary search by defining $a_1 = 1$ and $b_1 = 4$.
    For all $j$, we define 
    \[
    	\GuessedOPT_j \ \coloneqq \ \frac{2 a_j + b_j}{3}
    	\qquad \text{and} \qquad 
    	\varepsilon_j \ \coloneqq \ \frac{b_j - a_j}{36} \enspace.
   	\]
   	Note that $\GuessedOPT_j \in I_j \subseteq [1,4]$.
    In iteration $j$ of the binary search, we apply \cref{cor:guaranteed_behaviour_nice} 
    with $\GuessedOPT = \GuessedOPT_j$ and $\eta = \varepsilon_j$.
    If the algorithm concludes that $\OPT > \GuessedOPT_j$,
    we define $a_{j+1} = \GuessedOPT_j = \frac{2 a_j + b_j}{3}  $ and $b_{j+1} = b_j$.
    Otherwise, we obtain a solution $\bar s\in X$ with 
    \begin{align*}
    	\orderednorm{\bar s} \ &\leq \ \GuessedOPT_j \cdot \left( 1 + 3 \varepsilon_j \right) \
    	\leq \ \GuessedOPT_j + 4 \cdot 3 \cdot \varepsilon_j \
    	= \  \frac{2 a_j + b_j}{3}  + \frac{b_j - a_j}{3} \
    	= \ \frac{a_j + 2 b_j}{3}   \enspace;
   	\end{align*}
    then we define $a_{j+1} = a_j$ and $b_{j+1} = \frac{a_j + 2 b_j}{3} \geq \OPT$.
    
    In either case, $b_{j+1} - a_{j+1} = \frac{2}{3} (b_j - a_j)$.
    Since $b_1 - a_1 = 3 \in \bigO{1}$, this second stage of binary search terminates after $k \in \bigO{\log \left( \frac{1}{\varepsilon} \right) }$ iterations.
    
    The  total number of gradient evaluations is dominated by the number of gradient evaluations in the 
    last call to \cref{cor:guaranteed_behaviour_nice}
    because the number of gradient evaluations in iteration $j$ is in $\bigO{\frac{n+d}{\varepsilon_j^2}\log(n+d)}$ and
    \[
    	\sum_{j=1}^k \frac{1}{\varepsilon_j^2} \
    	= \ \frac{1}{\varepsilon_k^2} \cdot \sum_{i=0}^{k-1} \left(\frac{4}{9}\right)^i \
    	\leq \ \frac{1}{\varepsilon_k^2} \cdot \frac{9}{5}  \enspace .
   	\]
    Therefore, the total number of gradient evaluations in the second stage of the binary search is in $\bigO{\frac{n+d}{\varepsilon^2}\log(n+d)}$.

The number of oracle calls normally obeys the same bound as the number of gradient evaluations.
However, we have $\bigO{\log\frac{1}{\varepsilon}+\log\log d}$ applications of \cref{cor:guaranteed_behaviour_nice},
and we need to take the union bound for the events that the number of oracle calls is larger than that bound.
To obtain a bound that is independent of $\varepsilon$, we call the algorithm of
\cref{cor:guaranteed_behaviour_nice} up to $2^{k-j}$ times in iteration $j$ of the second stage of binary search, 
and in each of these $2^{k-j}$, except the last one, we abort after $4\Omega$ oracle calls.
The probability that in one iteration $j$ we run over the $4\Omega$ oracle calls in the last attempt is 
less than $2(nd)^{-20\cdot 2^{k-j}}$, which is dominated by the last iteration and sums up to less than $3(nd)^{-20}$.
Adding a union bound for the first stage of the binary search, and using 
$3+\lceil\log_2\lceil \log_2 d \rceil \rceil \le d^2$, we conclude that the 
probability that we ever need more than $4\Omega$ oracle calls in any application of \cref{cor:guaranteed_behaviour_nice}
is at most $(nd)^{-18}$.

If none of these bad events occur (i.e., with probability at least $1-(nd)^{-18}$), the number
of oracle calls obeys the same bound as the number of gradient evaluations, which however is now
larger due to our multiple attempts. However, in the second stage of the binary search, these are still
dominated by the only attempt in the last iteration because 
    \[
    	\sum_{j=1}^k \frac{2^{k-j}}{\varepsilon_j^2} \
    	= \ \frac{1}{\varepsilon_k^2} \cdot \sum_{i=0}^{k-1} \left(\frac{8}{9}\right)^i \
    	\leq \ \frac{9}{\varepsilon_k^2}  \enspace . \qedhere
   	\]
\end{proof}

\subsection{Approximate Oracles} \label{section:approximate_oracles}

In many applications (such as fractional Steiner tree packing \cite{muller2011faster}), the linear minimization problem over $X_c$ is hard to solve,
but can be approximated efficiently up to a relative error of $\tau$ for some constant $\tau \geq 1$,
i.e., the oracle functions $f_c: \Rdpos \rightarrow X_c$ return solutions that satisfy
\[
\inp{f_c(y)}{y} \ \leq \ \tau \cdot \min \left\{ \inp{x}{y} : x \in X_c \right\} \qquad \text{for all } y \in \Rdpos \ , \ c \in C  \enspace .
\]
Clearly, under this assumption, no guarantee below $\tau \cdot \OPT$ is achievable.
Our algorithm and proof techniques are applicable to such a setting;
we obtain a $\tau (1+\varepsilon)$-approximation within the same asymptotic running time.

\cref{thm:continuous_approx,lemma:continuous_runtime,thm:continuous_runtime_deterministic} and their proofs remain unchanged.
In \cref{lemma:enoughprogress}, we need to change the assumption $\GuessedOPT \geq \OPT$ to $\GuessedOPT \geq \tau \cdot \OPT$; then the proof works the same way because
$\sum_{c \in A} \inp{\nabla \Psi(S(t))}{s_c} \le \tau \sum_{c\in A} \min_{x\in X_c} \inp{\nabla \Psi(S(t))}{x} \le \tau \cdot \OPT$.
In \cref{alg:main_algorithm_with_guaranteed_termination} (Lines 10 and 16) and
\cref{cor:guaranteed_behaviour,cor:guaranteed_behaviour_nice},
the conclusion $\GuessedOPT<\OPT$ needs to be replaced by $\GuessedOPT < \tau \cdot \OPT$.

In \cref{thm:main_theorem_precise}, we need to make some changes in the proof, adapting the binary search.

In the first stage of the binary search, we get the weaker bound of $\onenorm{v} \leq \tau \cdot d \cdot \OPT$.
Therefore, we create $\lceil \log_2(\tau d)\rceil$ scaled instances.
The first binary search will identify an index $i$ such that $1 \leq \OPT^{(i)} \leq 4 \tau$
by changing the parameter $\GuessedOPT=2$ to $\GuessedOPT = 2 \tau$ in the configuration.
If the algorithm terminates, we get a solution $x \in X$ with $\orderednorm{x} \leq 4 \tau$.
Otherwise, we can conclude $\OPT^{(i-1)} > 1$.
Since $\tau$ is a constant, the asymptotic running time stays the same.
After the first stage of the binary search, we can assume that we have an instance where $\OPT \in [1, 4 \tau]$.

In the second stage of the binary search, our sequence of intervals has the property that $a_j \leq \tau\cdot\OPT$.
It is not necessarily the case that $b_j \geq \tau \cdot \OPT$, but we always have a solution $x^{(j)}$ with $\orderednorm{x^{(j)}} \leq b_j$ by construction.
Thus, once $b_i - a_i \leq \varepsilon \tau$, we found a $\tau (1+\varepsilon)$-approximation.
To start, we define $a_1 = 1$ and $b_1 = 4 \tau$.
The definitions of $\GuessedOPT_j$ and $\varepsilon_j$ and the rest of the proof remain unchanged.

Putting everything together, this results in the following theorem for approximate linear minimization oracles:

\begin{theorem}
	\label{thm:approximate_oracles}
    Let $\orderednorm{\cdot}$ be an ordered norm on $\Rd$ with an $\left( \ln d, 1\right)$-approximation $\Psi$.
    Assuming we are given a linear minimization oracle with constant-factor approximation ratio $\tau \geq 1$,
    there is an algorithm that, for any given instance and any $\varepsilon \in (0,1]$,
    computes a solution $x \in X$ with $\orderednorm{x} \leq (1+\varepsilon) \tau \cdot \OPT$, 
    and makes at most $\Omega^*$ gradient evaluations of $\Psi$, at most $n\Omega^*$ oracle calls, 
    and more than $\Omega^*$ oracle calls with probability at most $(nd)^{-18}$. 
    \qed
\end{theorem}

The remaining piece to prove \cref{thm:main} and its generalization to approximate oracles is an $\left( \ln d, 1\right)$-approximation of any ordered norm. 

\section{Construction of an Optimal Norm Approximation}
\label{sec:construction_norm_approximation}

In the following, we show that an $\left( \frac{\ln d}{\eta}, \eta \right)$-approximation $\Psi_{\eta}$ of $\orderednorm{\cdot}$ is given through (a scaled version of) the convex conjugate of the \textit{negative entropy} $F$ defined on an appropriate dual space $\dualspace$.
For the $\ell_{\infty}$ norm, this norm approximation is the \textsc{LogSumExp} function, and the gradients are (proportional to) the well-known \emph{multiplicative weights}.
It is worth noting that the gradients of $\Psi_{\eta}$ can be interpreted as prices of the \textit{Follow-the-Regularized-Leader} (FTRL) algorithm
with $F$ as regularizer.
We highlight this connection by providing a regret based analysis of~\cref{alg:ordered_norm_minimization_singe_customer_continuous} in~\cref{subsec:single_cust_ftrl_analysis}.

The goal of this section is to prove~\cref{thm:norm_approximation}, which, in addition to the existence of such an approximation, states that its gradients can be computed efficiently.
To prove~\cref{thm:norm_approximation}, we need a few definitions first.

\paragraph{Dual Space}
We can write every ordered norm as
\[\orderednorm{x} \ = \ \inp{\sort{x}}{\beta} \ = \ \max \left\{\big. \inp{x}{y}: y \text{ is a permutation of } \beta \right\} \ = \ \max \left\{\big. \inp{x}{y}: y \in Y \right\}\]
for
\begin{align*}
	\dualspace \
	&\coloneqq \ \conv \left(\big. \left\{ y :  y \text{ is a permutation of } \beta \right\} \big.\right)
	\\[.2cm]
	&= \ \left\{ y \in \Rdpos : \ \sum_{i \in J} y_i \leq \sum_{i \in J} \beta_i \ \text{ for all} \ J \subseteq \{1,\dots,d\} \ , \ \sum_{i=1}^d y_i = \sum_{i=1}^d \beta_i \right\}
	\enspace .
\end{align*}
Therefore, we call $\dualspace$ the \emph{dual space} that corresponds to $\orderednorm{\cdot}$.
Note that  $\dualspace$ contains all the non-dominated vectors with dual norm equal to one, i.e., all $y \in \Rdpos$ with $\onenorm{y}=1$ and $\dualorderednorm{y}=1$.

\paragraph{Negative Entropy}
The \emph{negative entropy} $F:\dualspace \to \mathbb{R}$ on $\dualspace$ is defined by
\[
	F(y) \ \coloneqq  \ \sum_{i=1}^d y_i \ln y_i \qquad \text{for} \ y \in \dualspace \enspace,
\]
defining $y_i \ln y_i = 0$ for $y_i=0$.
Its convex conjugate $F^*:\Rdpos \to \mathbb{R}$ is defined by
\[
	F^*(x) \ \coloneqq \ \max_{y \in \dualspace} \left( \inp{x}{y} - F(y) \right) \qquad \text{for } x \in \Rdpos \enspace .
\]
It is well-known (see e.g.~\cite{rockafellar1997convex}) that the gradient of the convex conjugate is given by the maximizing argument of the right hand side,
\begin{equation}
	\nabla F^*(x) \ = \ \arg \max_{y \in \dualspace} \left( \inp{x}{y} - F(y) \right) \qquad \text{for } x \in \Rdpos \enspace .
	\label{eq:convex_conjugate_gradient}
\end{equation}

\paragraph{Generalized Relative Entropy Projection}
Let $D_F$ denote the \emph{Bregman divergence}~\cite{bregman1967relaxation} induced by $F$, i.e.,
\[
	D_F (q,p) \ \coloneqq \ F(q) - F(p) - \inp{\nabla F(p)}{q-p} \qquad \text{for } p,q \in \Rdpos \enspace ,
\]
and, for $p \in \Rdpos$, define the \emph{generalized relative entropy projection} of $p$ to $\dualspace$ by
\[
	\Proj_{\dualspace}(p) \
	\coloneqq \ \arg \min_{y \in \dualspace} \ D_{F}(y,p) \
	= \ \arg \min_{y \in \dualspace} \
	\sum_{i=1}^d \left( y_i \ln \left( \frac{y_i}{p_i} \right) - y_i + p_i \right) \
	= \ \arg \min_{y \in \dualspace} \
	\sum_{i=1}^d y_i \ln \left( \frac{y_i}{p_i} \right).
\]
The last equation holds because $\sum_{i=1}^d y_i = 1$ for all $y \in \dualspace$.
Due to strict convexity of $F$, $\Proj_{\dualspace}(p)$ is unique.

The following relation between the gradient of the convex conjugate and the projection to $Y$ is also well-known.
\begin{proposition}
	\label{prop:ftrl_prices}
    For all $x \in \Rdpos$,
	\begin{equation}
		\label{eq:gradients_as_projection}
		\nabla F^*(x) \ = \ \Proj_{\dualspace} \bigl( \exp(x_1), \dots, \exp(x_d) \bigr) \enspace .
	\end{equation}
\end{proposition}

\begin{proof}
	This follows immediately from~\eqref{eq:convex_conjugate_gradient}, since
	\begin{align*}
		\Proj_{\dualspace}(\exp(x_1), \dots, \exp(x_d)) \ & = \ \arg \min_{y \in \dualspace} \
		\sum_{i=1}^d y_i \ln \left( \frac{y_i}{\exp(x_i)} \right)\\[.2cm]
		& = \ \arg \min_{y \in \dualspace} \left( F(y) - \inp{y}{x} \right) \\[.2cm]
		& = \ \nabla F^*(x) \enspace. \qedhere
	\end{align*}
\end{proof}

For $\eta=1$, the convex conjugate $F^*$ will be our norm approximation.
To prove~\cref{thm:norm_approximation}, we need to prove three properties about $F^*$.
First, we need to show that $F^*$ approximates $\orderednorm{\cdot}$ within an additive error.
Second, we need to show that $\nabla F^*$ is stable in the sense of~\cref{def:norm_approximation}, and third, that $\nabla F^*$ can be computed efficiently.
The characterization of $\nabla F^*$ provided by~\cref{prop:ftrl_prices} is central to prove the second and the third property.
The first property can be shown easily as follows.

\begin{lemma}
	\label{lem:norm_apx_ftrl}
    For all $x \in \Rdpos$,
	\[\orderednorm{x} \ \leq \ F^*(x) \ \leq \ \orderednorm{x} + \ln d \enspace.\]
\end{lemma}

\begin{proof}
	Let $x \in \Rdpos$.
	First, by non-positivity of $F$ on $\dualspace$,
	$F^*(x) \geq \max_{y \in \dualspace} \inp{x}{y} = \orderednorm{x}$.
	Second,
	\[
		F^*(x) \
		\leq \ \max_{y \in \dualspace} \inp{x}{y} \ + \ \max_{y \in \dualspace} \left( -F(y) \right) \ = \ \orderednorm{x} \ + \  \max_{y \in \dualspace} \left( -F(y) \right)  \enspace.
	\]
	Using that the maximum of $-F(y)$ over the probability simplex $\probsimplex$ is attained at $y = \frac{1}{d} 1^d$ with $-F \left( \frac{1}{d}1^d \right) = \ln d$, and that we have $\frac{1}{d} 1^d \in \dualspace \subseteq \probsimplex$, the above yields
	\[
		\orderednorm{x} \ \leq \ F^*(x) \ \leq \ \orderednorm{x} + \ln d  \qquad \text{ for all } x \in \Rdpos \enspace .\qedhere
	\]
\end{proof}
The second property, i.e., gradient stability of $F^*$, is implied by the following contraction property of the projection.
We defer the proof of this property to~\cref{sec:proof_contraction_property}.
\begin{restatable}[Contraction property]{theorem}{contractionproperty}
	\label{thm:contraction_property}
	Let $ \xi \geq 1$ and $p, q \in \Rdstrictlypos$ with $q \leq p \leq \xi q$.
	Then,
	\begin{equation}
		\Proj_{\dualspace}(p) \ \leq \ \xi \cdot \Proj_{\dualspace}(q).
		\label{eq:contraction_projection}
	\end{equation}
\end{restatable}

Given~\cref{lem:norm_apx_ftrl} and~\cref{thm:contraction_property}, we prove that $F^*$ yields our desired norm approximation:

\begin{theorem}
	\label{thm:norm_approximation_convex_conjugate}
	Let $\eta > 0$.
	Then $\Psi_{\eta}(x) = \frac{1}{\eta} F^*(\eta x)$ is an $\left( \frac{\ln d}{\eta}, \eta\right)$-approximation of $\orderednorm{\cdot}$.
\end{theorem}
\begin{proof}
	Let $x \in \Rdpos$.
	First, by~\cref{lem:norm_apx_ftrl},
	\[
		\orderednorm{x} \ = \ \frac{\orderednorm{\eta x}}{\eta} \ \leq \ \frac{ F^*(\eta x)}{\eta} \ = \ \Psi_{\eta}(x) \ \leq \ \frac{1}{\eta} \cdot \left( \orderednorm{\eta x} + \ln d\right) \ = \ \orderednorm{x} + \frac{\ln d}{\eta} \enspace.
	\]
	Second, consider $x,s \in \Rdpos$.
	According to~\cref{prop:ftrl_prices} and~\cref{thm:contraction_property}, we have
	\begin{align*}
		\nabla \Psi_{\eta}(x+s) \ &= \ \nabla F^*(\eta (x+s)) \\[.1cm]
		&= \ \Proj_{\dualspace} \bigl( \exp(\eta (s_1 + x_1)), \dots, \exp(\eta (s_n + x_d)) \bigr)\\[.1cm]
		&\leq \ \exp \bigl( \eta \inftynorm{s} \bigr) \cdot \Proj_{\dualspace}(\exp(\eta x_1), \dots, \exp(\eta x_d)) \\[.1cm]
		&= \ \exp \bigl( \eta \inftynorm{s} \bigr) \cdot \nabla F^*(\eta x) \\[.1cm]
		&= \ \exp \bigl( \eta \inftynorm{s} \bigr) \cdot \nabla \Psi_{\eta}(x) \enspace.
	\end{align*}
	Therefore, $\Psi_{\eta}$ is an $\left(\frac{\ln d}{\eta}, \eta\right)$-approximation of $\orderednorm{\cdot}$.
\end{proof}

Lastly, we need to show that $\nabla F^*$ can be computed efficiently.
According to~\cref{prop:ftrl_prices}, this is equivalent to the efficient computation of a generalized relative entropy projection to $Y$.
In~\cite{lim2016efficient}, generalized relative entropy projections to $\dualspace$ were studied in a different context (they call $\dualspace$ a generalized permutahedron).
They provide an $\bigO{ d \log d }$ algorithm to compute the projection.

\begin{theorem}[\cite{lim2016efficient}]
	\label{thm:projection_efficient_computation}
	For any given $p \in \Rdpos$,
	the generalized relative entropy projection $\Proj_{\dualspace}(p)$ can be computed in $\bigO{d \log d}$ time. \qed
\end{theorem}

\cref{thm:norm_approximation_convex_conjugate,thm:projection_efficient_computation,prop:ftrl_prices} imply~\cref{thm:norm_approximation}.
\cref{thm:norm_approximation} is best possible in the following sense:

\begin{proposition} \label{prop:bestpossiblenormapx}
There exists no $(o(\log d), 1)$-approximation of $\ell_{\infty}$.
\end{proposition}

\begin{proof}
	The special case of fractional packing with constant width $\rho > 0$ can be described as finding $\min_{x \in X} \inftynorm{x}$
	for $X \subseteq [0,\rho]^d$ (and $n = 1$) in our setting.
	In~\cite{klein2015number}, it was shown that any (even randomized) algorithm of the Dantzig--Wolfe type, that is,
	an algorithm iteratively querying linear minimization oracles and returning a convex combination of the oracle solutions,
	needs $\Omega\left(\varepsilon^{-2} \log d \right)$ oracle calls to compute a solution $x \in X$ of value $\inftynorm{x} \leq 1+\varepsilon$, assuming $\OPT = 1$, $\rho \geq 2$ and $d^{- \frac{1}{4}} (\ln d) < \varepsilon < \frac{1}{10} $.

	Let $\alpha \in o(\log d)$.
	Assume we are given an $(\alpha, 1)$-approximation of $\ell_{\infty}$, which, by scaling, gives also rise to an $\bigl(\frac{\alpha}{\eta}, \eta\bigr)$-approximation of $\ell_{\infty}$ for any $\eta > 0$.
	The approximation ratio of \cref{alg:ordered_norm_minimization_singe_customer_continuous} in~\cref{lem:guarantee_single_cust} after $T \geq \frac{\ln d}{\eta^2}$ iterations assumed that an $\bigl(\frac{\ln d}{\eta}, \eta\bigr)$-approximation was given.
	It is easy to see that the same approximation guarantee holds after $T \geq \frac{\alpha}{\eta^2}$ iterations when an $\bigl(\frac{\alpha}{\eta}, \eta\bigr)$-approximation is used.
	More precisely, \cref{alg:ordered_norm_minimization_singe_customer_continuous} computes a solution $\bar{s}$ with $\orderednorm{\bar{s}} \leq \eta + \frac{\exp(\eta) - 1}{\eta} \OPT$ after $T \geq \frac{\alpha}{\eta^2}$ oracle calls given an $\bigl(\frac{\alpha}{\eta}, \eta\bigr)$-approximation of $\ell_{\infty}$, assuming $X \subseteq [0,1]^d$.
	By scaling the instance with $\frac{1}{\rho}$, this can be used to compute a solution of value at most $(1+\varepsilon)\OPT$ after $T \in \bigO{\varepsilon^{-2} \cdot \rho \cdot \alpha} = \bigO{\varepsilon^{-2} \cdot \alpha}$ oracle calls.

	But since~\cref{alg:ordered_norm_minimization_singe_customer_continuous} is of the Dantzig--Wolfe type, this is not possible.
\end{proof}

\section{Proof of the Contraction Property}
\label{sec:proof_contraction_property}

In this section, we prove~\cref{thm:contraction_property}.
To do so, we first need to derive structural properties of the projected point.
In~\cite{suehiro2012online}, the projection to $\dualspace$ was studied in a different context 
(they describe $\dualspace$ as the base polyhedron of a cardinality-based submodular function),
and several structural properties of the projected point were provided.
We revisit these structural properties in the following, as they are crucial for the proof of~\cref{thm:contraction_property}.

Let $p \in \Rdpos$.
For $j=0,\ldots,d$, we write $[j]=\{1,\ldots,j\}$ and use the notation 
$B_j = \sum_{i\in[j]} \beta_i$ and $P_j = \sum_{i\in[j]} p_i$ for the partial sums of $\beta$ and $p$ in the following.
Note that $B_d = 1$.
The projection $\Proj_Y(p)$ is given as the optimal solution to
\begin{equation}
	\begin{aligned}
		\text{minimize}   \qquad & D_F(y,p) \\[.1cm]
		\text{subject to} \qquad & \sum_{i \in I} y_i \le B_{|I|} \qquad \forall I \subseteq [d] \\[.1cm]
		& \sum_{i=1}^d y_i = 1 \\[.1cm]
		& \qquad y \geq 0 \enspace .
	\end{aligned}
	\label{eq:bregman_projection_all_subsets}
\end{equation}
An important observation is that when $p$ is sorted, problem~\eqref{eq:bregman_projection_all_subsets} is equivalent to the following optimization problem with only $d$ constraints.
This was stated without a complete proof in~\cite{suehiro2012online}.

\begin{equation}
	\begin{aligned}
		\text{minimize}   \qquad & D_F(y,p) \\[.1cm]
		\text{subject to} \qquad & \sum_{i=1}^j y_i \le B_j \qquad \forall j=1,\dots, d-1 \\[.1cm]
		& \sum_{i=1}^d y_i = 1 \\[.1cm]
		& \qquad y \geq 0
	\end{aligned}
	\label{eq:bregman_projection_base_polyhedron}
\end{equation}
We say that constraint $j \in [d]$ is \emph{tight}, if $\sum_{i=1}^j y_i = B_j$.

Clearly, every feasible solution to~\eqref{eq:bregman_projection_all_subsets} is a feasible solution to~\eqref{eq:bregman_projection_base_polyhedron}.
Further, every sorted solution to~\eqref{eq:bregman_projection_base_polyhedron} is a feasible solution to~\eqref{eq:bregman_projection_all_subsets}.
The following lemma states that every optimal solution to~\eqref{eq:bregman_projection_base_polyhedron} is sorted, establishing equivalence of~\eqref{eq:bregman_projection_all_subsets} and~\eqref{eq:bregman_projection_base_polyhedron} for sorted vectors $p$.

\begin{lemma}
	\label{lem:optimal_solutions_are_sorted}
	Assume that $p \in \Rdpos$ is sorted in non-increasing order.
	Then, the optimal solution $y$ to~\eqref{eq:bregman_projection_base_polyhedron} is also sorted in non-increasing order.
	Therefore, the optimum solutions to \eqref{eq:bregman_projection_all_subsets} and \eqref{eq:bregman_projection_base_polyhedron} are the same.
\end{lemma}

\begin{proof}
	The generalized relative entropy $D_F(y,p) = s(y) + l(y)$ can be decomposed into a strictly convex symmetric part $s(y) = \sum_{i=1}^d (y_i \ln (y_i) - y_i + p_i)$ and a linear part $l(y) = \sum_{i=1}^d y_i (- \ln p_i)$.
	Since $p$ is sorted in non-increasing order, the coefficients of $l$ are non-decreasing in the index $i$.

	Now assume that there is an optimum solution $y$ to~\eqref{eq:bregman_projection_base_polyhedron} that is not sorted in non-increasing order.
	Let $h \in [d-1]$ be an index where consecutive indices are increasing, i.e., $y_h < y_{h+1}$.
	We define $y^{(1)} = y$ and $y^{(2)}$ as the vector that arises form $y$ by swapping the entries $y_{h}$ and $y_{h+1}$.
	We claim that $\tilde{y} = \frac{1}{2} \left( y^{(1)}+y^{(2)} \right)$ is a feasible solution to~\eqref{eq:bregman_projection_base_polyhedron}, and that $D_F(\tilde{y}, p) < D_F(y,p)$, contradicting the optimality of $y$.

	First, by symmetry of $s$, $s \left( y^{(1)} \right) = s \left( y^{(2)} \right)$, which, by strict convexity of $s$, leads to 
	\[
	s(\tilde{y}) \ < \ \frac{1}{2} \bigl( s \bigl( y^{(1)} \bigr) + s \bigl( y^{(2)} \bigr) \bigr) \ = \ s(y) \enspace .
	\]
	Second, since the coefficients of $l$ are non-decreasing, $l \left( y^{(2)} \right) \leq l \left( y^{(1)} \right)$, and therefore, 
	\[
	l(\tilde{y}) \ = \ \frac{1}{2} \bigl( l \bigl(y^{(1)}\bigr) + l\bigl(y^{(2)}\bigr)\bigr) \ \leq \ l(y) \enspace .
	\]
	These two observations imply $D_F(\tilde{y}, p) < D_F(y,p)$.

	It remains to show feasibility of $\tilde{y}$ to~\eqref{eq:bregman_projection_base_polyhedron}.
	Since the partial sums of $\tilde{y}$ correspond to the partial sums of $y$ for all indices $j \in [d]\setminus\{h\}$, and $y$ is a feasible solution,
	we only need to verify the constraint $\sum_{i=1}^h \tilde{y}_i \leq B_h$.
	Let $\tau = B_{h+1} - \sum_{i=1}^{h-1} y_i \geq y_h+y_{h+1}$.
	Note that (by defining $B_0 = 0$)
	\[
	\tau \ = \ \underbrace{B_{h-1} \ - \ \sum_{i=1}^{h-1}y_i}_{\geq 0, \text{ by feasibility of $y$}} \ + \ \beta_h \ + \ \beta_{h+1} \enspace .
	\]
	Since $\beta_h \geq \beta_{h+1}$,
	\[
	\frac{\tau}{2} \ \leq \ B_{h-1} \ - \ \sum_{i=1}^{h-1}y_i \ + \ \beta_h \ = \ B_h \ - \ \sum_{i=1}^{h-1}y_i \enspace .
	\]
	Combining our observations, we can conclude feasibility of $\tilde{y}$ by
	\[
	\sum_{i=1}^h\tilde{y}_i  \ = \ \sum_{i=1}^{h-1} y_i  \ + \ \frac{1}{2}(y_h + y_{h+1}) \ \leq \ \sum_{i=1}^{h-1} y_i \ + \ \frac{\tau}{2} \ \leq \ B_h \enspace .  
	\]

We have shown that every optimal solution to~\eqref{eq:bregman_projection_base_polyhedron} is sorted.
Note that any sorted solution to~\eqref{eq:bregman_projection_base_polyhedron} is a feasible solution to~\eqref{eq:bregman_projection_all_subsets}.
On the other hand, the constraints of~\eqref{eq:bregman_projection_all_subsets} are a superset of the constraints of~\eqref{eq:bregman_projection_base_polyhedron}; hence, any solution to~\eqref{eq:bregman_projection_all_subsets} is a feasible solution to~\eqref{eq:bregman_projection_base_polyhedron}. 
This proves that the optimum solutions to \eqref{eq:bregman_projection_all_subsets} and \eqref{eq:bregman_projection_base_polyhedron} are the same.
\end{proof}

Structural properties of the projected point heavily rely on quotients of partial sums of $\beta$ and $p$.
Therefore, we introduce the following notation.
For $0\le i < j \le d$, we denote the \emph{interval ratio} associated with these two indices by
\[
\intervalratio_p(i,j) \ \coloneqq \ \frac{B_j - B_i}{P_j - P_i} \enspace .
\]
The following lemma summarizes some simple calculation rules for interval ratios.

\begin{lemma}
	Let $i,j,l \in [d]$ with $i < j < l$ and $p,q \in \Rdstrictlypos$ with $q \leq p$. Then:
	\begin{enumerate}[label=(\roman*)]
		\item $\sum_{h=i+1}^j p_h \intervalratio_p(i,j) \ = \ B_j - B_i \enspace .$
		\item $\intervalratio_p(i,j) \ \leq \ \intervalratio_q(i,j) \enspace .$
		\item $\intervalratio_p(i,j) \ \leq \ \intervalratio_p(j,l)$ if and only if $\intervalratio_p(i,j) \ \leq \ \intervalratio_p(i,l) \enspace .$
	\end{enumerate}
	\label{lem:calculations_with_interval_ratios}
\end{lemma}
\begin{proof}
	$(i)$ and $(ii)$ are trivial.
	$(iii)$ follows because $\intervalratio_p(i,l)$ is a strictly convex combination of $\intervalratio_p(i,j)$ and $\intervalratio_p(j,l)$.
	Indeed, $\intervalratio_p(i,l) = \alpha \intervalratio_p(i,j) + (1-\alpha) \intervalratio_p(j,l)$ for $\alpha = \frac{P_j - P_i}{P_l - P_i}$ and $0<\alpha<1$.
\end{proof}

We say that the ratios $\frac{y_i}{p_i}$ for $i \in [d-1]$ change at a constraint $j$ if $\frac{y_j}{p_j} \neq \frac{y_{j+1}}{p_{j+1}}$.
Note that if constraints $i,j$ with $i < j$ are tight and $\alpha = \frac{y_{i+1}}{p_{i+1}} = \dots = \frac{y_{j}}{p_j}$, then $\alpha = \intervalratio(i,j)$.
Thus, if the ratios change only at tight constraints, there is a partition of $[d]$ into intervals that provide the ratios $\frac{y_i}{p_i}$ through their interval ratios.
This is one of the central structural properties of the projected point.
The following theorem provides useful equivalent descriptions of the projected point.

\begin{theorem}[Characterization of the projected point]
	\label{thm:characterization_projected_point}
	Let $p \in \Rdstrictlypos$ with entries sorted in non-increasing order.
	Let $y \in \Rdpos$.
	The following conditions are equivalent to $y = \Proj_{\dualspace}(p)$.
	\begin{enumerate}[label=(\roman*)]
		\item $y$ is an optimal solution to~\eqref{eq:bregman_projection_all_subsets}.
		\item $y$ is an optimal solution to~\eqref{eq:bregman_projection_base_polyhedron}.
		\item $y$ is a feasible solution to~\eqref{eq:bregman_projection_base_polyhedron}, the ratios $\alpha_i = \frac{y_i}{p_i}$ are non-decreasing in $i$, and $\alpha_i \neq \alpha_{i+1}$ implies $\sum_{j=1}^i y_j = B_i$.
		\item $y$ is a feasible solution to~\eqref{eq:bregman_projection_base_polyhedron}, and there is a $k \in \mathbb{N}$ and indices $0 = i_0 < i_1 < \dots < i_k =d$ such that for all $j=0, \dots, k-1$ and $i \in \{i_{j}+1,\dots, i_{j+1}\}$,
		\[
		\frac{y_i}{p_i} \ \overset{(a)}{=} \ \intervalratio_p(i_j, i_{j+1}) \ \overset{(b)}{=} \ \min_{l > i_j} \ \intervalratio_p(i_j, l) \enspace . 
		\]
	\end{enumerate}
\end{theorem}

These conditions were mostly provided in~\cite{suehiro2012online}.
They state (ii) without a complete proof, (iii) follows from the KKT conditions, and (iv) is not stated explicitly in~\cite{suehiro2012online}, but essentially follows from the correctness of their algorithm.
Nonetheless, we provide a self-contained proof here.

\begin{proof}
	Note that (i) describes the projection of $p$ by definition, and (i) and (ii) are equivalent by \cref{lem:optimal_solutions_are_sorted}.

	\medskip
	\noindent\textbf{(ii) \bm{$\Leftrightarrow$} (iii).}
	Observe that $y>0$ for every optimal solution $y$ to~\eqref{eq:bregman_projection_base_polyhedron}.
	This is because for every feasible solution $y$ with an entry $i \in [d]$ with $y_i=0$, we can find another entry $j \in [d]$ such that increasing $y_i$ and decreasing $y_j$ by some small $\varepsilon > 0$ will result in another feasible solution $y'$. 
	Now for sufficiently small $\varepsilon$, this solution $y'$ has smaller divergence $D_F(y',p) < D_F(y,p)$, 
	because $\nabla_{y_j} D_F(y',p)= \ln ( \frac{y_j-\epsilon}{p_j} ) >  \ln ( \frac{\epsilon}{p_i} ) = \nabla_{y_i} D_F(y',p)$
	for all $0<\epsilon<\frac{y_jp_i}{p_i+p_j}$.
		
	Through the KKT conditions, combined with $y>0$, (ii) is equivalent to the following three conditions.
	First, $y$ is a feasible solution to~\eqref{eq:bregman_projection_base_polyhedron}.
	Second, there are $\mu_1, \dots, \mu_d \geq 0$ and $\lambda \in \mathbb{R}$,
	such that $-\nabla D(y,p) = \sum_{i=1}^d \mu_i  a_i + \lambda 1^d$, where $a_i = (1^i, 0^{d-i})$ is the coefficient vector of the $i$-th constraint.
	Third, for all $j\in [d]$, complementary slackness holds: $\mu_j \left( \sum_{i=1}^j y_i - B_j \right) = 0$.
	The second condition is equivalent to the existence of $\lambda \in \mathbb{R}, \mu_1, \dots, \mu_d \geq 0$,
	such that
	\begin{equation}
		\ln \left( \frac{p_j}{y_j} \right) \ = \ \lambda + \sum_{i=j}^d \mu_i \qquad \text{for all } j \in [d] \enspace .
		\label{eq:ordered_norms_projection_kkt}
	\end{equation}
	This is equivalent to the ratios $\frac{p_j}{y_j}$ being non-increasing, or equivalently, the ratios $\frac{y_j}{p_j}$ being non-decreasing.
	By complementary slackness, $\mu_j >0$ implies tightness of the corresponding constraint, i.e., $\sum_{i=1}^j y_i = B_j$.
	Thus, by~\eqref{eq:ordered_norms_projection_kkt}, the ratios $\frac{y_j}{p_j}$ may only change at tight constraints.

	\medskip
	\noindent\textbf{(iii) \bm{$\Rightarrow$} (iv).}
	Assume (iii).
	Since the ratios $\frac{y_i}{ p_i}$ only change at tight constraints, there must be indices $0 = i_0 < i_1 < \dots < i_k =d$ such that $(a)$ is fulfilled.
	To show that $(b)$ is also fulfilled, we observe that by feasibility and the non-decreasing order of the ratios for any $i_j$ and $l > i_j$,
	\begin{equation}
		B_l \ \geq \ \sum_{i=1}^l y_i \ = \ \sum_{i=1}^l \frac{y_i}{p_i} \cdot p_i \ = \ B_{i_j} + \sum_{i=i_j +1}^l \frac{y_i}{p_i} \cdot p_i \ \geq \ B_{i_j} + \frac{y_{i_j + 1}}{p_{i_j +1}} \cdot \left( P_{l} - P_{i_j}\right) \enspace ,
		\label{eq:ratio_feasibility}
	\end{equation}
	i.e., $\intervalratio_p(i_j, i_{j+1}) = \frac{y_{i_j+1}}{p_{i_j + 1}} \leq \min_{l > i_j} \intervalratio_p(i_j, l)$.

	\medskip
	\noindent\textbf{(iv) \bm{$\Rightarrow$} (iii).}
	Assume (iv).
	Note that $\sum_{i=i_j+1}^{i_{j+1}}y_i = \sum_{i=i_j+1}^{i_{j+1}}p_i \intervalratio_p(i_j, i_{j+1})$.
	Therefore, using~\cref{lem:calculations_with_interval_ratios} (i), we conclude that the constraints $i_1, \dots, i_k$ are tight,
	implying that the ratios $\frac{y_j}{p_j}$ change only at tight constraints.
	Further, by assumption, $\intervalratio_p(i_j, i_{j+1}) \leq \intervalratio_p(i_j, i_{j+2})$, implying $\intervalratio_p(i_j, i_{j+1}) \leq \intervalratio_p(i_{j+1}, i_{j+2})$ according to \cref{lem:calculations_with_interval_ratios} (iii),
	showing that the ratios are non-decreasing.
\end{proof}

Finally, we have the tools to prove~\cref{thm:contraction_property}.

\contractionproperty*

\begin{proof}
	Let $y\coloneqq \Proj_{\dualspace}(p) $ and $z \coloneqq \Proj_{\dualspace}(q)$.
	We assume that the entries of $p$ are sorted in non-increasing order; this is no loss of generality because $Y$ is symmetric and $D_F$ is separable.
	Denote the ratios between projected and unprojected points by $\alpha \coloneqq \left(\frac{y_1}{p_1}, \dots, \frac{y_d}{p_d} \right)$ and $\gamma \coloneqq \left( \frac{z_1}{q_1}, \dots, \frac{z_d}{q_d} \right)$.

	\medskip
	\noindent\textbf{Case 1: \bm{$q$} is sorted in non-increasing order.}
	We prove $\alpha \leq \gamma$, which implies the claim for this case.
	Assume $\alpha \not\leq \gamma$
	and let $i \in [d]$ be the smallest index such that $\alpha_i > \gamma_i$.
	According to~\cref{thm:characterization_projected_point} (iv), we have $\alpha_i = \intervalratio_p(l_p, u_p)$ and
	$\gamma_i = \intervalratio_q(l_q, u_q)$ for some $l_p,l_q < i \leq u_p, u_q$.
	Since $q$ is sorted, $\gamma$ is non-decreasing by \cref{thm:characterization_projected_point}~(iii).
	It must hold that $l_p = i-1$, because otherwise we would have $\alpha_{i-1} = \alpha_{i} > \gamma_i \geq \gamma_{i-1}$,
	and $i$ would not be the smallest index with $\alpha_i > \gamma_i$.
	We have
	\[ 
	\alpha_i \ = \ \intervalratio_p(i-1, u_p) \ \overset{\text{\cref{thm:characterization_projected_point} (iv) (b)}}{\leq} \ \intervalratio_p(i-1, u_q) \
	\overset{\text{\cref{lem:calculations_with_interval_ratios} (ii)}}{\leq} \ \intervalratio_q(i-1, u_q) \enspace .
	\]
	If $l_q = i-1$, this immediately leads to $\alpha_i\le \intervalratio_q(i-1, u_q)=\gamma_i$, a contradiction.
	Otherwise, $l_q < i-1$ and $\intervalratio_q(l_q, u_q) \leq \intervalratio_q(l_q, i-1)$, again by \cref{thm:characterization_projected_point} (iv) (b).
	By~\cref{lem:calculations_with_interval_ratios} (iii), this implies $\gamma_i = \intervalratio_q(l_q, u_q) \geq \intervalratio_q(i-1, u_q) \geq \alpha_i$,
	leading to a contradiction.

	\medskip
	\noindent\textbf{Case 2: \bm{$q$} is not sorted in non-increasing order.}
	Note that, by symmetry, $\Proj_{\dualspace}(\sort{q}) =  \sort{z}$.
	Further, the facts that $p$ is sorted and 
	$q\le p\le \xi q$ easily imply $\sort{q} \leq p \leq \xi \sort{q}$ by comparison of rearrangements.
	Therefore, for $i \in [d]$ with $z_i \geq \left( \sort{z} \right)_i$, we get from the first part of the proof (applied to $\sort{q}$)
	that $\frac{y_i}{z_i} \leq \frac{y_i}{(\sort{z})_i} \leq \xi$.
	For $i \in [d]$ with $z_i < \left( \sort{z} \right)_i$, the first part of the proof (again applied to $\sort{q}$) yields  
	$\alpha \leq \left( \frac{(\sort{z})_1}{(\sort{q})_1}, \dots,  \frac{(\sort{z})_d}{(\sort{q})_d} \right)$. 
	The right-hand side is a permutation of $\gamma$ due to \cref{lem:optimal_solutions_are_sorted}, and it is non-decreasing by 
	\cref{thm:characterization_projected_point}~(iii); so $\alpha\leq \gamma^{\uparrow}$.
	Further, $z_i < \left( \sort{z} \right)_i$ implies $\gamma_i \geq \left( \gamma^{\uparrow} \right)_i$, again by \cref{thm:characterization_projected_point}~(iii).
	Together, $\frac{z_i}{ q_i} = \gamma_i \geq \left( \gamma^{\uparrow} \right)_i  \geq \alpha_i = \frac{y_i}{p_i}$, which implies
	$\frac{y_i}{z_i} \leq \frac{p_i}{q_i} \leq \xi$.
\end{proof}

\newpage

\bibliography{paper}
\bibliographystyle{plain}

\newpage

\appendix

\section{Relation to Follow-the-Regularized-Leader}
\label{subsec:single_cust_ftrl_analysis}

Instead of using the gradients of a norm approximation $\Psi$ as prices to determine the next solution~$s^{(t)}$ with our minimization oracle over $X$ (cf.\ \cref{sec:simple_algorithm}), 
we could also compute these prices explicitly, based on the current sum of solutions $S(t)$, as done in \textit{Follow-the-Regularized-Leader} (cf.\ \cite{Lattimore_Szepesvari_2020}),
i.e., $y^{(t)} = \Proj_{\dualspace} \left( \exp \left( \eta \cdot S(t) \right) \right)$, and
$s^{(t+1)} = \arg \min_{s \in X} \inp{y^{(t)}}{s}$ for $t \in \{0, \dots, T\}$.
Note that for the norm approximation derived in \cref{lem:norm_apx_ftrl}, this yields the same algorithm and final solution
(cf.\ \cref{prop:ftrl_prices}).

To analyze the version with \textit{Follow-the-Regularized-Leader}, we can utilize the standard \emph{regret} bound, with $F$ as regularizer and over $\dualspace$,
\begin{align*}
	R_T \
	&\coloneqq \ \max_{y \in \dualspace} \ \sum_{t=1}^T  \inp{y - y^{(t-1)}}{s^{(t)}} \\[.2cm]
	&\stackrel{\text{\small \cite{Lattimore_Szepesvari_2020}}}{\leq} \ \max_{y \in \dualspace} \ \frac{1}{\eta} \left( F(y) - F\left( y^{(0)} \right) \right) \ + \ \sum_{t=1}^T \inp{y^{(t)}-y^{(t-1)}}{s^{(t)}} \ - \ \frac{1}{\eta} \ \sum_{t=1}^T D_F \left( y^{(t)}, y^{(t-1)} \right) \\[.2cm]
	&\leq \ \frac{- F\left( y^{(0)} \right)}{\eta} \ + \ \sum_{t=1}^T \inp{y^{(t)}-y^{(t-1)}}{s^{(t)}}
	\enspace ,
\end{align*}
and, by weak duality,
\[
	R_T \ = \ \orderednorm{S(T)} - \sum_{t=1}^T \inp{y^{(t-1)}}{s^{(t)}} \
	\stackrel{ \eqref{eq:weak_duality}}{\geq} \ T \cdot \left(\orderednorm{\bar{s}} - \OPT \right) \enspace .
\] 
This also implies an additive error of $\eta$ for $T \geq \eta^{-2} \ln d$ by $-F \left( y^{(0)} \right) = \max_{y \in \dualspace} - F(y) = \ln d$.
The contraction property of the projection (cf.\ \cref{thm:contraction_property}) now replaces bounded gradient increase of the norm approximation and leads to a multiplicative error of $\exp(\eta)$.
This is slightly worse than the guarantee in \cref{lem:guarantee_single_cust}.
For $\eta > 0$, we get
\[
	\orderednorm{\bar{s}} \ \leq \ \eta + \exp(\eta) \cdot \OPT \enspace ,
\]
which implies $\orderednorm{\bar{s}} \leq \eta + (1+2\eta) \cdot \OPT$ for $\eta \leq 1$.

\section{Related Work}
\label{sec:related_work}
\label{sec:previous_work_and_limitations}
The fractional load balancing problem lies at the intersection of convex optimization, combinatorial optimization, and online algorithms.
It generalizes classical problems such as fractional packing and maximum concurrent flow, while simultaneously differing from these settings in two crucial ways:
first, we assume access to the feasible sets $X_c$ only via linear minimization oracles; second, whereas these classical problems are formulated as $\ell_{\infty}$-minimization problems, we consider more general classes of norms.

Each of the areas mentioned above provides algorithmic techniques that address certain aspects of the problem.
However, none of the existing approaches simultaneously handle general norms, work with oracle access to the customer sets, and achieve a number of oracle calls that is near-linear in the number of customers and resources.
This section reviews the most relevant approaches from these areas and explains why they do not directly yield the guarantees required in our setting.

A central structural feature of fractional load balancing is the presence of multiple customers.
The set of feasible solutions decomposes as a Minkowski sum $X = \sum_{c \in C} X_c$,
where $X_c$ denotes the set of feasible load vectors for customer $c$.
In the terminology of convex optimization, this induces a so-called \emph{block structure} in which each customer corresponds to one block.

\subsection{Perspectives from Convex Optimization}

\paragraph{Projection-Based First-Order Methods.}

Minimizing a norm $\norm{\cdot}$ over a convex set $X$ is a fundamental task in convex optimization.
A standard textbook approach is \emph{projected subgradient descent}, which computes a solution with additive error $\varepsilon$ in
$\bigO{D^2\varepsilon^{-2}}$ iterations, where $D$ denotes the diameter of $X$ with respect to $\norm{\cdot}$; see, e.g., \cite[Chapter~3]{boyd2004convex}.
Each iteration requires a subgradient evaluation of $\norm{\cdot}$ and a projection to $X$.

In our setting, we assume access to $X$ only through a linear minimization oracle.
Although projection to $X$ can in principle be implemented via the equivalence of separation and optimization together with the ellipsoid method~\cite[Chapter~6]{grotschel2012geometric}, this approach incurs substantial overhead and is therefore not compatible with the running times targeted in our work.
Consequently, classical projection-based first-order methods are not well suited to our model.

A standard technique for addressing non-smooth objectives is \emph{smoothing}.
Observe that the norm minimization problem can be formulated as a \emph{bilinear min-max problem}
\begin{equation}
	\min_{x \in X} \norm{x}
	=
	\min_{x \in X} \max_{y \in \Rd, \dualnorm{y} \le 1} \inp{x}{y} , \label{eq:saddle_point_formulation}
\end{equation}
where $\dualnorm{\cdot}$ denotes the dual norm of $\norm{\cdot}$.
This structure enables the use of smoothing and primal-dual first-order methods for structured saddle-point problems, including \emph{Nesterov's smoothing} and \emph{excessive gap} techniques~\cite{nesterov2005smooth,nesterov2005excessive} and \emph{Nemirovski's mirror prox} method~\cite{nemirovski2004prox}.
These methods build on the broader mirror descent framework originating in the work of Nemirovski and Yudin~\cite{nemirovskij1983problem}.
For suitably smoothed objectives, they achieve a number of iterations of order $\bigO{\varepsilon^{-1}}$ to obtain an additive error $\varepsilon$, where the running time additionally depends on geometric parameters of the feasible region.

However, these methods rely on efficient so-called proximal (or mirror) steps over the primal set of feasible solutions $X$, which in particular require projections to $X$.
As discussed above, such operations are not directly available.

Nevertheless, the min-max perspective is conceptually related to our approach.
In \cref{sec:construction_norm_approximation}, we constructed stable norm approximations by regularizing the min-max formulation with the negative entropy over a suitably chosen dual space $Y$.

\paragraph{Conditional Gradient Methods.}

Given that efficient projections to $X$ are not available in our oracle model, a natural alternative is the \emph{conditional gradient} method (also known as the \emph{Frank--Wolfe} method)~\cite{frank1956algorithm},
which minimizes a differentiable convex function $\Psi$ over a compact convex set using only a linear minimization oracle.
For smooth convex objective functions $\Psi$, the classical Frank--Wolfe algorithm achieves an additive error of order $\varepsilon$ using $\bigO{\varepsilon^{-1}}$ oracle calls, where the constant depends on a curvature constant of $\Psi$ over $X$ \cite{jaggi2013revisiting}.
This curvature constant can be upper bounded by $L \cdot D^2$, where $L$ is the Lipschitz constant of $\nabla \Psi$ and $D$ is the diameter of $X$.
When the objective is a norm such as $\ell_{\infty}$, a common approach is to replace it by a smooth approximation, for example, the \textsc{LogSumExp} function.
Since this approximation incurs an additive error of order $\Theta(\varepsilon)$ while the gradient Lipschitz constant scales as $L=\Theta \left( \varepsilon^{-1}\log d \right)$,
this yields a bound on the number of iterations of $\Omega \left( D^2(\log d)\,\varepsilon^{-2} \right)$, which is insufficient when $X$ has large diameter.

Numerous refinements of the classical Frank--Wolfe method have been proposed.
Variants such as \emph{away-step} and \emph{pairwise} Frank--Wolfe algorithms~\cite{lacoste2015global} and \emph{blended} or \emph{accelerated} conditional gradient methods~\cite{braun2025conditional,braun2025blendedconditionalgradientsunconditioning} achieve faster convergence under additional assumptions,
for example, when the feasible region is a polytope with favorable geometric properties or when the objective satisfies stronger curvature or strong convexity properties.
However, these improvements rely on geometric properties of $X$ or special structures of the objective function $\Psi$ that are not available in our general setting.

In the multi-customer setting, a natural idea is to apply \emph{block-coordinate} variants of the conditional gradient method that update only a single block in each iteration.
Such algorithms typically select blocks either cyclically~\cite{beck2015cyclic,sun2015improved} or uniformly at random~\cite{lacoste2013block}.
However, as discussed in \cref{sec:simplealganditslimitations}, algorithms that treat all customers symmetrically require $\Omega(n\cdot d)$ oracle calls.
Thus, although block-coordinate conditional gradient methods have been studied in the literature, they do not by themselves yield the running time guarantees targeted in our work.

\subsection{Perspectives from Online Algorithms}

The bilinear min-max formulation~\eqref{eq:saddle_point_formulation}
naturally connects fractional load balancing to online learning and online convex optimization.
This formulation can be interpreted as a two-player zero-sum game.
The \emph{primal player} chooses a feasible load vector $x\in X$, representing the total load induced by the customers, while the \emph{dual player} chooses a vector $y$ from the dual unit ball, which can be interpreted as a price or penalty assigned to the resources.
The payoff of the game is $\inp{x}{y}$, which measures the weighted load under the current price vector.

From this perspective, computing a minimum-norm load vector corresponds to finding an equilibrium of this zero-sum game.
If the dual player chooses price vectors $y$ using an online learning algorithm while the primal player responds by minimizing $\inp{x}{y}$ over $x\in X$, then standard regret guarantees imply that the average iterates converge to an approximate equilibrium of the game, that is, a primal load vector with minimum norm together with a corresponding optimal dual vector.
Such \emph{no-regret} algorithms for online linear optimization include \emph{Mirror Descent}, \emph{Follow-the-Perturbed-Leader}, and \emph{Follow-the-Regularized-Leader} (FTRL).
These methods are closely related and can often be viewed as different interpretations of the same algorithm; see, e.g., \cite{shalev2012online,abernethy2014online,mcmahan2011follow,bach2015duality}.

In particular, the Follow-the-Regularized-Leader perspective is conceptually close to our approach.
The gradients of our smooth approximations play the role of price vectors and admit an interpretation as FTRL updates (see \cref{subsec:single_cust_ftrl_analysis}).
Thus, our algorithm can be viewed equivalently as operating in the primal by minimizing a smooth norm approximation, or in the dual via regularized online learning.
Related approaches based on smooth norm approximations for online load balancing problems appear in \cite{kesselheim2022online,molinaro2017online}.
Online load balancing problems with regard to $\ell_p$ norms~\cite{azar2004all, molinaro2017online} or general monotone symmetric norms have been studied in \cite{kesselheim2022online, kesselheim2024supermodular}.
These approaches are not directly applicable to our setting, since they only yield poly-logarithmic approximation guarantees.
However,~\cite{kesselheim2022online, molinaro2017online} worked with notions of "good" norm approximations, which we draw inspiration from.

The multi-customer structure introduces an additional difficulty.
While the primal best response decomposes across customers, a straightforward implementation would require querying all customer oracles in every iteration.
There is a line of work on structured online convex optimization that exploits decompositions of the decision space into blocks or product domains, including \emph{block-coordinate mirror descent} and related methods~\cite{dang2015stochastic,shalev2013stochastic}.
Such algorithms typically update only a subset of blocks in each iteration and obtain regret guarantees for block-structured decision spaces.
However, they typically rely on cyclic or uniformly random block-selection rules or assume projection access to the blocks.
Consequently, as discussed earlier, they do not directly overcome the $\Omega(n\cdot d)$ barrier in our setting where only linear minimization oracles are available.

\subsection{Perspectives from Discrete Optimization}

\paragraph{Fractional Packing, Flows, and Resource Sharing.}

Many classical problems in combinatorial optimization can be formulated as minimizing the $\ell_\infty$ norm of a load vector.
A well-studied example is fractional packing, which asks to compute
\[
	\max_{x \in \mathbb{R}_{\geq 0 }^l}\ \{\langle 1^l,x\rangle : Ax \le 1^d\} \ ,
\]
for a nonnegative matrix $A \in \mathbb{R}_{\geq 0}^{d \times l}$.
By scaling the solution, this problem can equivalently be written as minimizing $\inftynorm{Ax}$ over the simplex $\Delta^l$, and therefore fits into our framework as a special case (consider $X = \{Ax : x \in \Delta^l\}$).

Algorithms for fractional packing and related min-max load balancing problems are often based on \emph{multiplicative weight updates} and exponential potential functions.
Early work by Plotkin, Shmoys and Tardos~\cite{plotkin1995fast} introduced a general Lagrangian-relaxation framework for fractional packing and covering problems that forms the basis of many multiplicative-weight algorithms.
Subsequent work developed diameter-independent algorithms~\cite{young2001sequential, khandekar2004lagrangian}.
Later, more specialized algorithms that operate explicitly on the matrix $A$ and achieve running times depending on $\varepsilon^{-1}$ rather than $\varepsilon^{-2}$ were developed~\cite{bienstock2006approximating,allen2019nearly}.

A parallel line of work focused on the maximum concurrent flow problem, which naturally exhibits a multi-customer structure.
Given a network with edge capacities and multiple source-sink pairs with demands (called \emph{commodities}), the goal is to route all commodities while minimizing the maximum load on any edge.
This objective can be viewed as minimizing the $\ell_\infty$ norm of the vector of edge loads, where customers correspond to the source-sink pairs and linear minimization oracles can be implemented by shortest path computations.
Algorithms based on multiplicative weights and exponential potential functions have led to efficient approximation schemes for this problem and its variants
\cite{shahrokhi1990maximum,garg2007faster,fleischer2000approximating,karakostas2008faster}.
These techniques have also had a significant impact on VLSI global routing \cite{muller2011faster,held2017global}.

A particularly influential improvement is the algorithm of Garg and Könemann~\cite{garg2007faster}, which introduced a non-uniform step size rule that significantly accelerates multiplicative-weight methods.
For the maximum concurrent flow problem with $n$ commodities and $d$ edges, this approach yields a $(1+\varepsilon)$-approximate solution using $\bigOtilde{ (n+d) \varepsilon^{-2} }$ shortest path computations.
To the best of our knowledge, this was the first algorithm for this special case of our problem that avoids $\Omega(n \cdot d)$ oracle calls.

While multiplicative-weight methods dominated early work on multicommodity flow, more recent algorithms use accelerated convex optimization techniques.
Starting with work by Sherman~\cite{sherman2017generalized,sherman2017area}, subsequent algorithms based on \emph{area-convex regularization} achieve running times of order $\bigOtilde{ dn \varepsilon^{-1} }$ for graphs with $d$ edges and $n$ commodities.
Importantly, this is a bound on the total running time, not on the number of oracle calls, and is therefore significantly faster than the previously described oracle-based methods, which require $\Omega \left( (d+n)\varepsilon^{-2} \right)$ shortest path computations.
Subsequent work refined these ideas and extended them to directed graphs~\cite{chen2025accelerated}.
These algorithms rely heavily on structural properties of flow problems, such as \emph{congestion approximators} and specialized linear-algebraic primitives for graphs.

The algorithms described above that achieve a linear dependence on $\varepsilon^{-1}$ for fractional packing or maximum concurrent flow rely crucially on the constraint matrix $A$ being explicitly available.
When the constraint matrix is not explicitly accessible and only a linear minimization oracle is available, Klein and Young~\cite{klein2015number} show lower bounds for \emph{Dantzig--Wolfe-type} algorithms; in particular, one cannot beat $\Omega \left( \varepsilon^{-2}\log d \right)$ oracle calls in general.

A different line of work focuses on capturing the multi-customer structure explicitly.
The \emph{min-max resource sharing} framework of Grigoriadis and Khachiyan~\cite{grigoriadis1994fast,grigoriadis1996coordination} considers multiple customers that contribute to a shared resource usage vector and aims to minimize the maximum load,
which corresponds exactly to our problem when the objective is the $\ell_{\infty}$ norm.
Their algorithms maintain exponential potentials based on the \textsc{LogSumExp} function, which is a smooth approximation of the $\ell_\infty$ norm.
Later work by Müller, Radke and Vygen~\cite{muller2011faster} significantly improved the running time for this model, obtaining a bound of $\bigOtilde { (n+d)\varepsilon^{-2} }$ oracle calls for instances with $n$ customers and $d$ resources.

Our results extend this type of guarantee beyond the $\ell_\infty$ objective: we obtain running times of order $\bigOtilde{ (n+d)\varepsilon^{-2} }$ for general ordered norms.
While proof techniques for the $\ell_{\infty}$ objective exploit the explicit structure of the \textsc{LogSumExp} function as well as a round-robin type analysis,
our algorithm for general ordered norms and its analysis are significantly more involved.

\paragraph{Integral Load Balancing and Ordered Norms.}
A related line of work studies integral load balancing problems and the minimization of ordered norms in combinatorial settings
\cite{chakrabarty2019simpler,chakrabarty2019approximation,ibrahimpur2021minimum,gupta2025balancing}.
In these problems, indivisible jobs or demands must be assigned to minimize objectives such as the maximum load, the sum of the largest $k$ loads, or more general ordered norms.
Algorithms in this area typically rely on combinatorial arguments, LP relaxations, and rounding techniques.

Since these integral problems are usually NP-hard, obtained approximation algorithms usually yield constant-factor guarantees.
Such guarantees permit relatively coarse approximations of the underlying norm and therefore tolerate larger additive errors in the norm approximation used as the objective function.
In contrast, our goal is to obtain $(1+\varepsilon)$-approximations, which requires different algorithmic techniques as well as norm approximations with small absolute error.

\subsection{Generalized Relative Entropy Projection}

In 1967 \cite{bregman1967relaxation}, Bregman introduced \textit{Bregman divergences} as a distance measure that relies on a strictly convex regularizing function, and studied projection under this divergence.
One prominent example in statistics and information theory is based on the (scaled) negative entropy $F$, resulting in the generalized relative entropy $D_F$ \cite{kruithof1937telefoonverkeersrekening,sinkhorn1964relationship,butnariu1997iterative,benamou2015iterative,kostic2023method,kostic2022batch}.
This is also the Bregman divergence that we use.
For probability distributions, it is known as
\textit{KL-divergence}, after Kullback and Leibler~\cite{kullback1951information}.

Our dual space $Y$, that we define over some fixed ordered norm $\orderednorm{\cdot}$ and that we want to project to, appears in the literature under different names,
such as the \emph{generalized permutahedron}~\cite{lim2016efficient} or the \emph{base polyhedron} described by the submodular function $f(I) = \sum_{i=1}^{|I|} (\sort{\beta})_i$ for $I \subseteq [d]$~\cite{suehiro2012online}.

For the special case of top-$k$ norms, a linear-time combinatorial projection algorithm was provided by Herbster and Warmuth \cite{herbster2001tracking} in the context of iterative online learning.

For the general case, Suehiro, Hatano, Kijima, Takimoto and Nagano \cite{suehiro2012online} established important structural properties of the projected point and developed an $\bigO{d^2}$ algorithm.
Lim and Wright \cite{lim2016efficient} attained an improved running time of $\bigO{d \log d}$ by reduction to the \emph{Isotonic Optimization Problem}~\cite{barlow1972isotonic},
for which efficient algorithms were developed independently before, e.g., the \emph{Pool Adjacent Violators} algorithm by Best, Chakravarti and Ubhaya \cite{best2000minimizing}, or the \emph{Single Row Clumping Algorithm} by Brenner and Vygen \cite{brenner2004legalizing} in the completely different context of VLSI placement.

\end{document}